\documentclass[a4paper,fleqn,usenatbib]{mnras}

\usepackage{mathptmx}

\usepackage[T1]{fontenc}
\usepackage{ae,aecompl}
\usepackage{tikz}
\usetikzlibrary{positioning}

\usepackage{graphicx}	% Including figure files
\usepackage{amsmath}	% Advanced maths commands
\usepackage{amssymb}	% Extra maths symbols

\title[Snow-lines can be thermally unstable]{Snow-lines can be thermally unstable}

\author[Owen, J. E.]{
James E. Owen\thanks{E-mail: james.owen@imperial.ac.uk}
\\
Astrophysics Group, Department of Physics, Imperial College London, Prince Consort Rd, London, SW7 2AZ, UK
}

\pubyear{2015}

\begin{document}
\label{firstpage}
\pagerange{\pageref{firstpage}--\pageref{lastpage}}
\maketitle

\begin{abstract}
Volatile species in protoplanetary discs can undergo a phase change from vapour to solid. These ``snow-lines'' can play vital roles in planet formation at all scales, from dust coagulation to planetary migration. In the outer regions of protoplanetary discs, the temperature profile is set by the absorption of reprocessed stellar light by the solids. Further, the temperature profile sets the distribution of solids through sublimation and condensation at various snow-lines. Hence the snow-line position depends on the temperature profile and vice-versa.  We show that this coupling can be thermally unstable, such that a patch of the disc at a snow-line will produce either run-away sublimation or condensation. This thermal instability arises at moderate optical depths, where heating by absorption of reprocessed stellar light from the disc's atmosphere is optically thick, yet cooling is optically thin. Since volatiles in the solid phase drift much faster than volatiles in the vapour phase, this thermal instability results in a limit-cycle. The snow-line progressively moves in, condensing volatiles, before receding, as the volatiles sublimate. Using numerical simulations, we study the evolution of the CO snow-line. We find the CO snow-line is thermally unstable under typical disc conditions and evolves inwards from $\sim50$ to $\sim30$~AU on timescales from 1,000-10,000 years. The CO snow-line spends between $\sim 10\%-50\%$ of its time at smaller separations, where the exact value is sensitive to the total opacity and turbulent viscosity. The evolving snow-line also creates ring-like structures in the solid distribution interior to the snow-line. Multiple ring-like structures created by moving snow-lines could potentially explain the sub-structures seen in many {\it ALMA} images.  
\end{abstract}

%%%%%%%%%%%%%%%%%%%%%%%%%%%%%%%%%%%%%%%%%%%%%%%%%%

%%%%%%%%%%%%%%%%% BODY OF PAPER %%%%%%%%%%%%%%%%%%

\section{Introduction}
Protoplanetary discs are composed of gas, which typically drives the dynamics, and solids, which can move with respect to the gas. Ultimately, it is the solids that are the building blocks from which planets form. The composition of the solids also varies strongly with position in the disc. As one moves to larger radii, volatile species (e.g. water, carbon dioxide, ammonia and carbon monoxide) are able to condense out of the vapour phase and into the solid phase, significantly enhancing the mass in the solids \citep[e.g.][]{Hayashi1981,Stevenson1988,Oberg2011,Booth2017}.

Since the condensation/sublimation of volatile species is a phase change, it happens over narrow ranges of temperature. In the context of protoplanetary discs this leads to rather abrupt changes in the solid density and composition, known as snow-lines (or ice-lines).  Snow-lines may play an important role in planet formation. Firstly, one does not just get an increase in solid surface density across the snow-line, but rather a strong local enhancement at the snow-line. This is because the strong concentration gradients allow volatiles in the vapour phase to diffuse back across the snow-line, recondensing into the solid phase, increasing the solid density locally at the snow-line \citep[e.g.][]{Stevenson1988,Cuzzi2004,Kretke2007}.  Secondly, the accretion of volatiles in either the vapour or solid phase will result in the sequestration of those elements, either in the planetary core or atmosphere. Therefore, understanding how volatiles are accreted by forming planets will allow us to interpret observations of the composition of exoplanet atmospheres \citep[e.g.][]{Oberg2011,Madhusudhan2014,Booth2017,Tinetti2018,Booth2019}.

The most commonly studied snow-line is the water snow-line. This is because water is one of the most abundant volatiles, resulting in a large solid density change \citep[e.g.][]{Hayashi1981,Stevenson1988}, and it is typically co-located between the terrestrial and giant-planets in our solar-system \citep[e.g.][]{Martin2012,Martin2013}. However, there are a multitude of other snow-lines, including but not limited to ammonia, nitrogen, carbon dioxide and carbon monoxide \citep[e.g.][]{Oberg2011,Martin2014,Booth2017}. 

The carbon monoxide (CO) snow-line is important in the astronomical context for several reasons. Firstly, CO is also an abundant volatile, resulting in large solid density changes at its snow-line \citep[e.g.][]{Oberg2011,Madhusudhan2014}. Secondly, CO is commonly used as a tracer of gas in protoplanetary discs through sub-mm observations. Thus, understanding CO's distribution is crucial to interpreting sub-mm fluxes and images of protoplanetary discs \citep[e.g.][]{Williams2014}. Finally, since the CO snow-line occurs in the outer regions of protoplanetary discs it (or its imprint) has been observed directly \citep[e.g.][]{Qi_science,Qi2019}. 

Understanding the location, evolution and chemistry of all snow-lines is going to be crucial to our understanding the chemistry of protoplanetary discs, planet formation and ultimately will allow us to interpret the compositions of exoplanet atmospheres. However, current work typically calculates the properties of snow-lines assuming the underlying disc temperature profile is fixed \citep[e.g.][]{Oberg2011,Booth2017}, or based on simple models that account for viscous heating \citep[e.g.][]{Kretke2007,Martin2012,Martin2014}. This approximation breaks down in protoplanetary discs, particularly in the outer regions. This is because it is the solid particles that absorb the stellar light and control the temperature of the disc. Therefore, changes in the distribution of solids will change the temperature profile which will change the distribution of solids. This is particularly important at snow-lines as small temperature changes can lead to large changes in the distribution of solids. 

In this work we show snow-lines, that occur at moderate optical depths in passively heated discs, are subject to thermal instabilities. Specifically, the snow-lines do not occur at fixed (or secularly evolving) locations as previously thought, but move about over large distances in otherwise stationary discs. This thermal instability fits within the picture of previously identified thermal instabilities in classical, active accretion discs, that are more commonly studied in high-energy astrophysics \citep[e.g.][]{Pringle1981,Faulkner1983,Lasota2001} or the earliest phases of star-formation \citep[e.g.][]{Armitage2001,Zhu2010,Martin2011,Martin2013}. The fact that many snow-lines are constantly evolving has important consequences for the chemistry of protoplanetary discs, the formation of planetesimals and planets and the interpretation of astronomical observations.

\section{Basic Concept} \label{sec:basic}
Thermal instabilities have been identified and studied in accretion disc physics since the origins of the field. Classical thermal instabilities arise at abrupt opacity changes or abrupt viscosity changes \citep[e.g.][]{Pringle1981}.  In these classical models, rapid changes in the thermal structure is associated with dramatic changes in accretion rates \citep{Lasota2001}. Within the standard $\alpha$ model, the kinematic viscosity is given by $\nu=\alpha_\nu c_s^2/\Omega$ (with $c_s$ the isothermal sound speed and $\Omega$ the Keplerian angular frequency). Since the mass-accretion rate is directly proportional to the kinematic viscosity, regions that are thermally unstable are also unstable to accretion rate variability. The outcome of such a thermal instability is accretion bursts followed by quiescence. 

Now clearly, such a scenario is not directly applicable to snow-lines. However, we can use the same language used to describe these classic secular thermal instabilities. Here we demonstrate the presence of a thermal instability and construct an ``S-curve'' to show the existence of a limit-cycle. 

\subsection{Qualitative description}

A thermal instability requires that the heating-rate is a stronger increasing function of temperature than the cooling rate. If this is the case, a system in thermodynamic equilibrium will be unstable to small temperature perturbations. For example, if the temperature increases slightly, then as the heating rate is a stronger increasing function of temperature, the small temperature increase results in a net heating rate, further increasing the temperature. Such an argument equally applies to a small temperature decrease, which leads to net cooling and continued reduction of the temperature. 

In the outer regions of a protoplanetary disc, the heating and cooling is almost exclusively set by radiative balance (i.e. viscous heating is negligible). Now consider a region of the disc with a temperature just below that required to sublimate a volatile. A small temperature increase will result in the volatile beginning to sublimate into the gas phase, decreasing the solid abundance. Since it is the solids that dominates the opacity of the disc, and hence the absorption and emission of radiation, the heating and cooling of the disc can be strongly affected by the decrease in solid abundance. 

In the outer regions of protoplanetary discs cooling is typically optically thin, hence this decrease in solid abundance will decrease the cooling rate. The heating of the disc arises from absorption of stellar irradiation re-radiated towards the disc mid-plane from the surface layers. Should the absorption of this re-radiated emission also be optically thin then the heating rate will also decrease in an equal fashion to the cooling rate and the disc will be thermally stable. However, should the absorption of this re-irradiated emission be at least marginally optically thick (or not very optically thin), then even though the solid abundance has decreased the disc will still absorb most of this re-radiated stellar light, hence the heating rate will remain unchanged. In this case, a small increase in disc temperature will result in drop in the cooling rate and a practically unchanged heating rate, leading to a net heating rate. Thus, the disc will increase in temperature leading to runaway sublimation of all the volatiles in the solid phase.   

The above argument also indicates a crucial requirement for this thermal instability: that the absorption of re-radiated stellar light and emission by the disc's mid-plane must occur at different optical depths. Namely, the absorption should be more optically thick than the cooling. Such a situation is natural in the outer regions of protoplanetary discs. Stellar irradiation of the surface layers of discs results in a ``super-heated'' surface-layer that is typically a factor of 4 hotter than the mid-plane \citep[e.g.][]{chiang97}. Since solid opacity increases with the increasing temperature of the radiation field, the optical depth to the re-irradiated stellar light is naturally larger than that of the emission from the disc's mid-plane. Hence, an important aspect of this instability when it comes to modelling is that it requires a non-grey treatment of not just the stellar radiation, but also the re-radiated stellar light from the disc's surface layers and the disc's mid-plane. If the disc is too optically thin, then the heating and cooling rate both respond linearly to the change in solid abundance and there is no instability. Further, if the disc is too optically thick the heating and cooling rates are essentially independent of the solid abundance\footnote{This ignores the fact that the fraction of stellar light absorbed in the surface layers does depend on the solid abundance; however, we ignore this effect in this simplified picture discussed here.}. Since the water snow-line occurs in regions of the disc that are optically thick under standard conditions (and where viscous heating is important -- \citealt{Kretke2007} and \citealt{Martin2011}), it is likely to be thermally stable. However, we speculate that snow-lines in the outer regions (e.g. CO, CO$_2$, NH$_3$) are thermally unstable.  

\subsection{Quantitative description}

In this sub-section we remain agnostic to the specific volatile species and discuss the requirements in general terms. A thermal instability in a disc requires:
\begin{equation}
    \frac{\partial \log Q_+}{\partial T_m} > \frac{\partial \log Q_-}{\partial T_m} \label{eqn:instability}
\end{equation}
where $Q_+$ and $Q_-$ and the heating and cooling rates of the disc per unit area respectively and $T_m$ is the mid-plane temperature. Taking the interior of the disc to be isothermal, with a temperature $T_m$, and containing a surface density of solids $\Sigma_d$, the cooling rate is  \citep[c.f.][]{Dullemond2001}:
\begin{equation}
    Q_-=2\sigma T_m^4 \frac{\int_0^\infty B_\nu(T_m) \left[1-\exp\left(-\Sigma_d\kappa_\nu/2\right)\right]{\rm d}\nu}{\int_0^\infty B_\nu(T_m) {\rm d}\nu} = 2 f_{\rm emit} \sigma T_m^4 \label{eqn:cool}
\end{equation}
where $B_\nu$ is the Planck function as a function of frequency ($\nu$), $\kappa_\nu$ is the solid opacity as a function of frequency, $\Sigma_d$ is the dust surface density and $f_{\rm emit}$ represents the cooling rate scaled to that of a pure black-body. The factor of 2 comes from the two sides of the disc. Further the heating rate is:
\begin{equation}
    Q_+ = F_{\rm surf} \frac{\int_0^\infty \kappa_\nu B_\nu(T_{\rm surf})  \left[1-\exp\left(-\Sigma_d\kappa_\nu/2\right)\right]{\rm d}\nu}{\int_0^\infty \kappa_\nu B_\nu(T_{\rm surf}) {\rm d}\nu} = f_{\rm abs} F_{\rm surf} \label{eqn:heat}
\end{equation}
where $F_{\rm surf}$ is the fraction of the stellar light re-radiated towards the mid-plane by the surface layers, $T_{\rm surf}$ is the temperature of the super-heated surface layers and $f_{\rm abs}$ represents the fraction of the re-radiated emission from the surface layers that is absorbed by the mid-plane.  

Now, in the optically thick limits $\Sigma_d \kappa \gg 1$, $f_{\rm emit}$ and $f_{\rm abs}$ are unity and there is no thermal instability (unless $\partial \log F_{\rm surf}/{\partial \log T_m} > 4$ which appears unlikely as the solid abundance decreases with increasing temperature at the snow-line), as expected from our qualitative discussion above.  

In the fully optically thin limit, e.g. $\Sigma_d \kappa \ll 1$ at all relevant frequencies, $f_{\rm emit}\rightarrow \Sigma_d \kappa_p(T_m)/2$ and $f_{\rm abs}\rightarrow \Sigma_d \kappa_p(T_s)/2$ \citep[e.g.][]{chiang97}, where $\kappa_p$ is the Planck mean opacity. As expected, both the heating and cooling rate linearly depend on $\Sigma_d$ in the optically thin limit. Thus, the only way to obtain a thermal instability is to have dramatic changes in the opacity due to the composition changes of the solid particles. While not implausible, the magnitude of the changes required $\partial \log \kappa_p(T_m) /\partial \log T_m \lesssim -4 $, means such a scenario may require a particularly special snow-line. 

However, at intermediate optical depths satisfying Equation~\ref{eqn:instability} becomes possible. To demonstrate this explicitly, while keeping our discussion agnostic about the species, we assume absorption and emission due to the solids can be described by geometric optics. Such that the opacity of a spherical grain, of size $s$, internal density ($\rho_i$), as a function of wavelength ($\lambda$) is given by:
\begin{equation}
    \kappa_\lambda (s) = \frac{3 Q(s,\lambda)}{4\rho_i s}
\end{equation}
where $Q(s,\lambda)$ is the absorption efficiency. In the geometric optics limit $Q=1$ when $\lambda < 2\pi s$, and $Q=(s/2\pi\lambda)^\beta$ when $\lambda > 2\pi s$, where $\beta$ is the Rayleigh index. 

\begin{figure}
    \centering
    \includegraphics[width=\columnwidth]{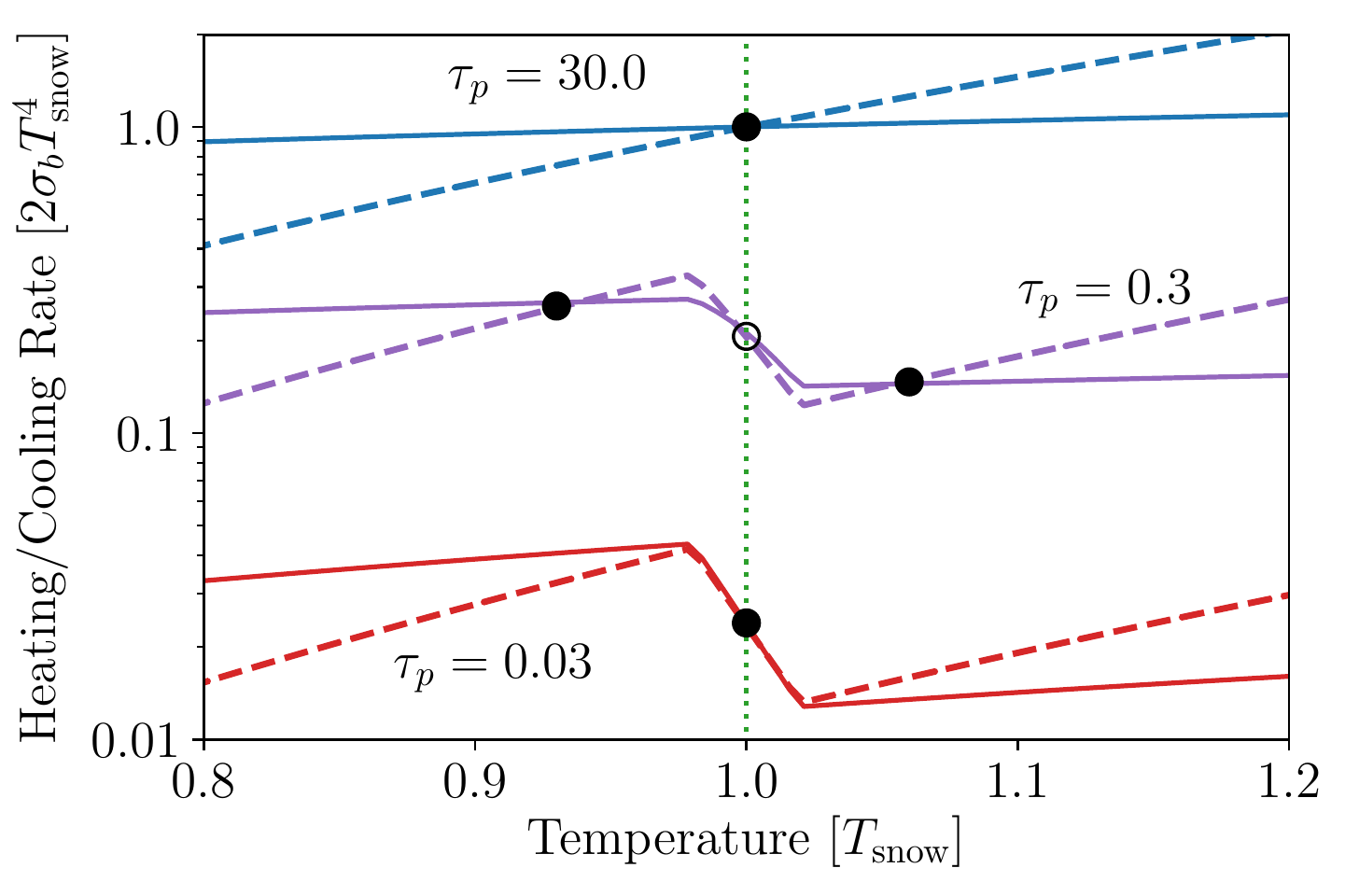}
    \caption{The heating (solid) and cooling rates (dashed) as a function of temperature in the vicinity of the snow-line. The lines are shown at Planck-mean optical depths with a temperature of $T_{\rm snow}$ for $\tau_p$ = 0.03, 0.3 \& 30. Filled circles show thermally stable equilibria, open circles show thermally unstable equilibria. For the heating rate curves, $F_{\rm surf}$ is taken to scale as $T^{1/2}$ (see Section~\ref{sec:model}), but is normalised such that thermal equilibrium is achieved at a temperature of $T_{\rm snow}$. At moderate optical depths the snow-line is thermally unstable, and thermally stable equilibria exist at both lower and higher temperatures. }
    \label{fig:QplusQminus}
\end{figure}

Now consider a scenario where the solid abundance drops from $\Sigma_d$ to $1/4\Sigma_d$ across the snow line at a temperature of $T_{\rm snow}$ with a width of 5\% of the snow-line temperature\footnote{Note we have chosen these parameters to emphasise the physical effect, rather than for a real snow-line, hence they are larger than typical snow-line values}. Calculations for physical parameters are shown in Section~\ref{sec:model}. The surface temperature $T_{\rm surf}= 4 T_{\rm snow}$. Figure~\ref{fig:QplusQminus} shows the heating and cooling rates as a function of temperature at a very high, moderate and very low optical depth. We choose a particle size that satisfies $2\pi s = b/T_{\rm ice}$ where $b$ is Wein's constant, use an internal density $1.25$~g~cm$^{-3}$ and adopt $\beta=1$.  This Figure shows (as discussed above) there is only one point at which heating balances cooling for the very high and very low optical depths all of which are thermally stable ($\partial \log Q_+/\partial T_m < \partial \log Q_-/ \partial T_m$). However, the moderate optical depth case shows three points at which heating balances cooling with the solution at a temperature of $T_{\rm snow}$ being thermally unstable ($\partial \log Q_+/\partial T_m > \partial \log Q_-/ \partial T_m$).

We note that provided the optical depths to cooling are moderate, the presence of a thermal instability is rather insensitive to the change in solid density, and how quickly this change happens, provided the transition occurs over a range of temperatures $\Delta T \ll T_{\rm snow}$. 

\subsection{Existence of a Limit Cycle}\label{sec:limit_simple}

One question we must now explore is how this thermal instability manifests in real protoplanetary discs. In the vicinity of the snow-line at moderate optical depths there are two possible temperature solutions at higher or lower temperatures than $T_{\rm snow}$, and further there must be a radius in the disc where its temperature ``jumps'' from the hot solution to the cold solution. However, a discontinuity in the temperature of the disc is not physical. Thus, by analogy with previously identified thermal instabilities in accretion discs it is likely to manifest itself as a limit-cycle, where the snow-line evolves in the disc. Therefore, a given radius in the disc will spend some of its time on the hot solution and some of its time on the cold solution.  

A limit-cycle occurs when there exists a forbidden range of accretion rates \citep[e.g.][]{Pringle1981}. In the context of our snow-line, we consider this to be a mass-flux of the volatile. This can either occur in the solid phase via radial drift, or in the gas-phase via advection along with the accreting gas. If the solids have grown sufficiently, such that radial drift is faster than gas advection, then a range of forbidden mass-fluxes of volatiles can exist. We will have a high volatile mass-flux when the disc is at the cold temperature solution as the volatiles are in the solid phase, and a low-mass flux when the disc exists at the hot temperature solution. If we feed the disc at a volatile mass flux that exists between the two allowed mass-fluxes then no steady-state can exist, and the disc will evolve through a limit-cycle. With the disc spending sometime in the cold, high accretion rate phase, and another fraction of the time in the hot, low accretion rate phase. 

\subsubsection{Construction of an ``S-curve''}

Consider a surface density of a given volatile $\Sigma_v$, if the volatile is in the solid phase the mass-flux is approximately \citep[e.g.][]{takeuchi02}:
\begin{equation}
    \dot{M}_v^{\rm ice} = 2\pi R \Sigma_v \tau_s  \left|\frac{{\rm d}\log P}{{\rm d}\log R}\right| \left(\frac{H}{R}\right)^2 v_K \label{eqn:ice_mdot}
\end{equation}
where $\tau_s$ is the Stokes number (or dimensionless stopping time)  of the solid particles, $H$ is the disc scale height and $P$ the mid-plane pressure. Whereas if the volatile is in the solid phase the mass-flux is approximately:
\begin{equation}
    \dot{M}_v^{\rm vapour} = 3\pi R \Sigma_v \alpha \left(\frac{H}{R}\right)^2 v_K \label{eqn:vapour_mdot}
\end{equation}
Therefore, comparing Equations~\ref{eqn:ice_mdot} \& \ref{eqn:vapour_mdot}, and noting the logarithmic pressure derivative is of order unity, we see the mass-flux in the ice-phase will be significantly higher than than in the vapour phase if the solids are undergoing radial drift (i.e. $\tau_s \gg \alpha$). Since this is the standard case for the outer regions of protoplanetary discs \citep[e.g.][]{birnstiel12}, we expect this to be the norm. 

\begin{figure}
    \centering
    \includegraphics[width=\columnwidth]{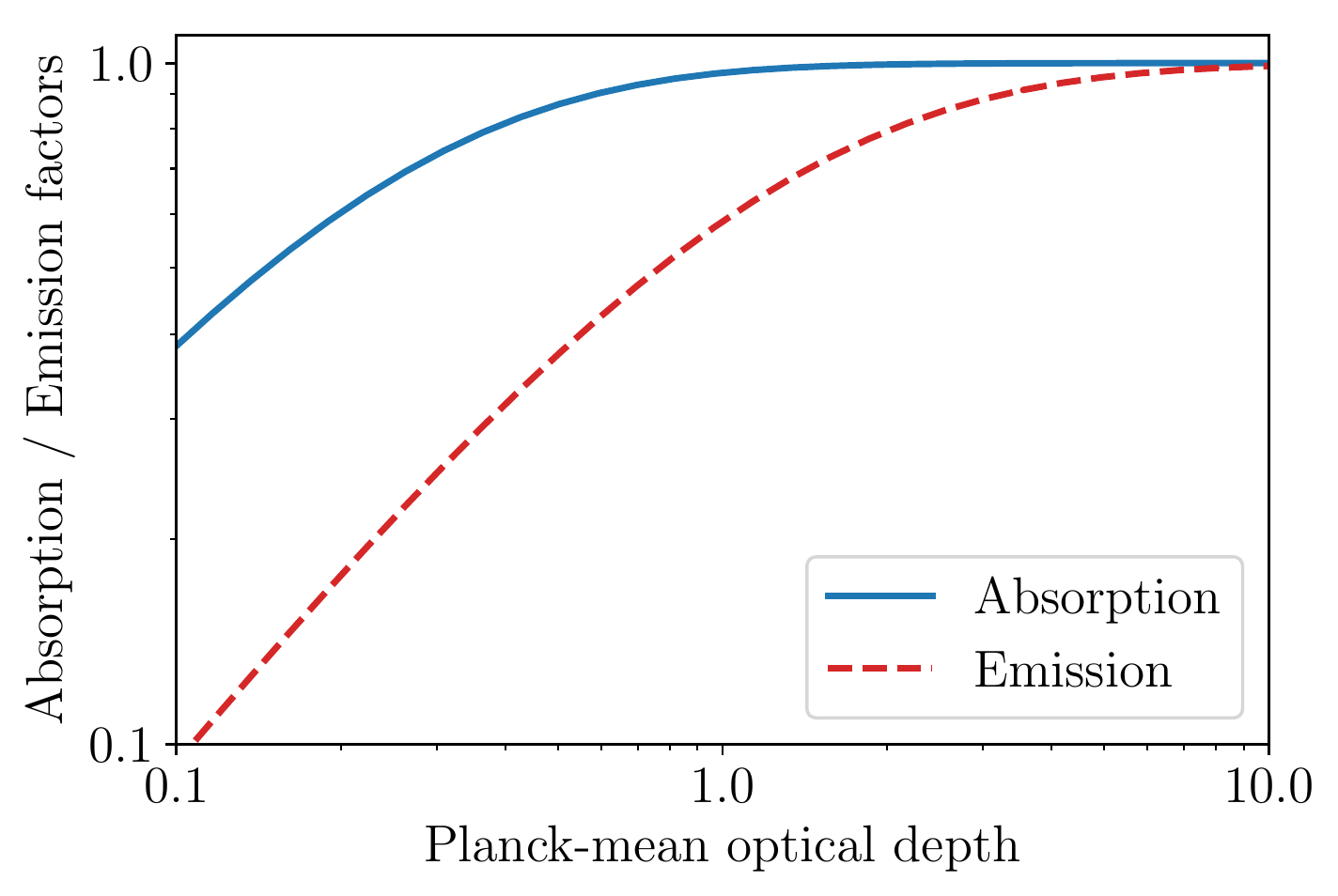}
    \caption{The absorption ($f_{\rm abs}$) and emission ($f_{\rm emit}$) efficiency factors as a function of Planck-mean optical depth. The curves are for $T_{\rm surf} = 4T_{\rm snow}$ and for particles whose optical properties are those of geometric optics. }
    \label{fig:f_factors}
\end{figure}

In order to understand how this gives rise to a limit-cycle it is useful to remind ourselves how $f_{\rm emit}$ and $f_{\rm abs}$ respond to changing the surface density. In the limit of moderate optical depth a reduction of the surface density means there is less material to emit so $f_{\rm emit}$ falls. Further, there is also less material to absorb the reprocessed stellar light so $f_{\rm abs}$ also falls. However, since opacity increases with the temperature of the radiation field, in the limit of moderate optical depth $f_{\rm emit}$ decreases more than $f_{\rm abs}$. This is clear if we plot $f_{\rm emit}$ and $f_{\rm abs}$ as a function of Planck-mean optical depth in Figure~\ref{fig:f_factors}, for our simple example discussed above (e.g. $T_{\rm surf}=4T_m$).

Now consider a region of the disc that is, initially, in the cold, high mass-flux state. Since we are feeding the disc with volatiles at a lower rate the surface density of volatiles will decrease, resulting in a lower solid surface density and optical depth. This decrease in optical depth will decrease both $f_{\rm emit}$ and $f_{\rm abs}$, but $f_{\rm emit}$ will decrease more (see Figure~\ref{fig:f_factors}), resulting in a warming of the disc. Eventually, the surface density of volatiles will drop so low that the disc will become hot enough to sublimate the volatiles into the vapour phase. This will dramatically decrease the mass-flux of the volatiles, to a rate lower than the rate at which the disc is being fed. Now the surface density of volatiles in this patch of the disc will increase, increasing the optical depth, which will cool the disc. Eventually the the disc will become so cool that the volatiles can fully condense into the solid phase, dramatically increasing the mass-flux of volatiles and the entire process can start over. 

\begin{figure}
    \centering
    \includegraphics[width=\columnwidth]{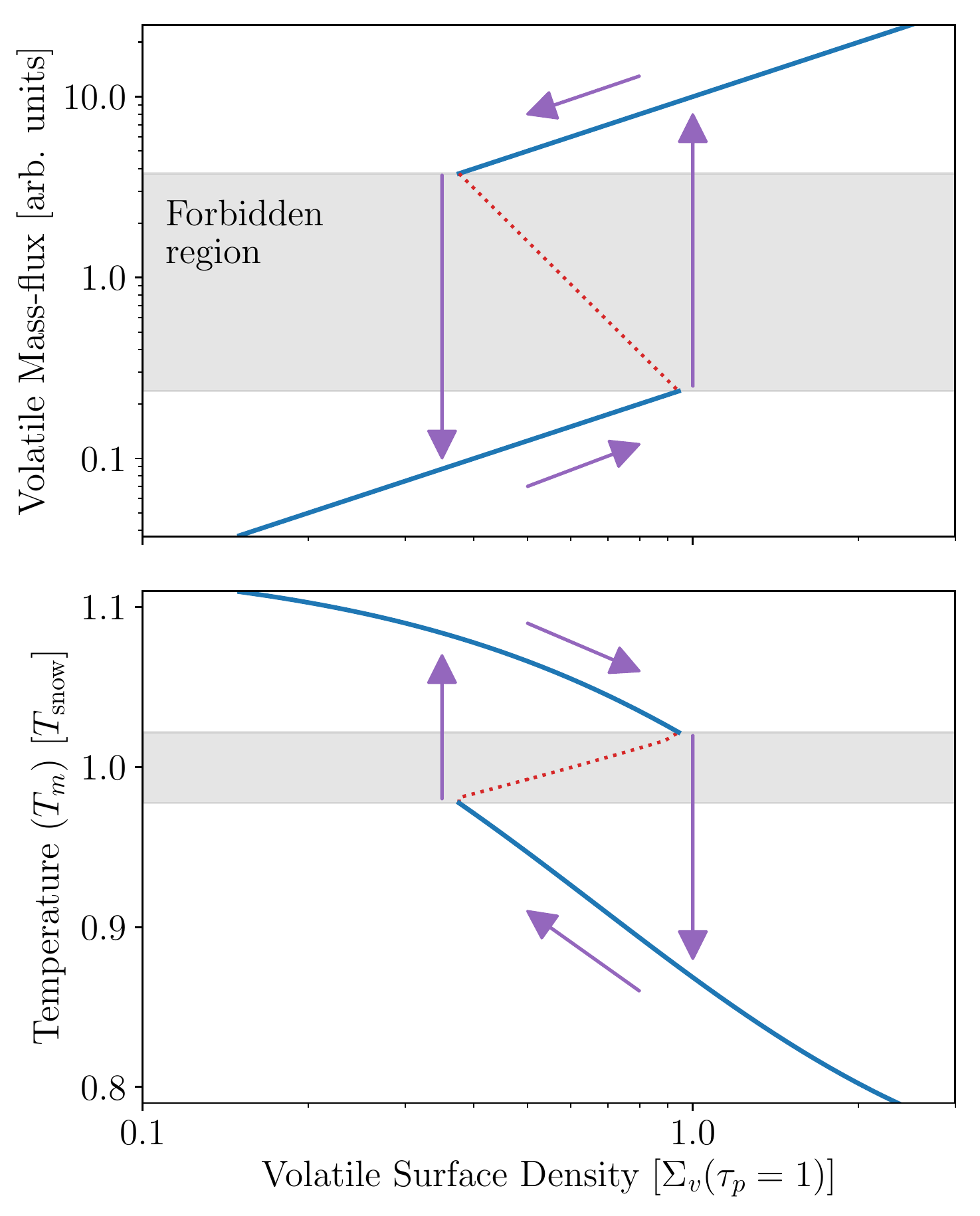}
    \caption{A sketch of a simple model for the limit-cycle. The top plot shows the S-curve of volatile surface density against volatile mass-flux. The bottom plot shows the disc temperature as a function of volatile surface density. The volatile surface density is scaled to a value to give a Planck optical depth of unity to radiation with a temperature of $T_{\rm snow}$. The solid blue lines represent thermally stable solutions, the dotted red lines represent thermally unstable solutions. No thermally stable states exist in the ``forbidden region'' of volatile mass-flux. The purple arrows sketch out the limit-cycle the disc's region will follow in both mass-flux and temperature. This schematic plot shows the limit cycle exists at moderate optical depths.  }
    \label{fig:simple_limit_cycle}
\end{figure}

The above limit-cycle can be pictorially represented by a ``S-curve'' where we plot the volatile mass-flux against the volatile surface density in Figure~\ref{fig:simple_limit_cycle}. This shows the evolution of both the surface density and temperature, with the arrows indicating the limit-cycle described above.  

\subsection{Complications to the simple picture}
The simple picture above motivated by previous S-curve analysis of thermal instabilities in discs is useful, and provides an intuitive insight to the presence of a thermal instability and a limit-cycle. However, unlike other classic thermal instabilities the heating rates and cooling rates and mass-fluxes cannot be expressed purely in local terms. For example, the ice-mass flux (Equation~\ref{eqn:ice_mdot}) is sensitive to the pressure derivative, which is sensitive to the temperature profile. As we shall see later, $F_{\rm surf}$ is also sensitive to the gradients of the temperature and density profile. Therefore, the above analysis is purely schematic. 

Finally, it is well known that radial transport of radiation and material across thermal fronts can have qualitative effects on the evolution of thermal instabilities in accretion discs \citep{Faulkner1983,Cannizzo1993,Owen2014}. In our snow-line case, the moderate optical depths give rise to short cooling times; thus, we don't expect advective or turbulent mixing of thermal energy to be important. However, the snow-line by definition contains very strong gradients in both solid and vapour concentrations and turbulent diffusion will play an important role in the evolution of the limit-cycle. Further, since the the outer regions of the snow-line are fed by radial drift -- which is sensitive to the temperature gradient -- the flux of solids (both icy and non-icy) is likely to evolve meaning our simplification of a fixed volatile mass flux is not quite correct. 

While all the above effects play a significant role in prescribing the exact evolution of the snow-line, the basic physics of why snow-lines are unstable and the existence of limit-cycles remains unchanged.

\section{Model - the CO snow-line} \label{sec:model}
In Section~\ref{sec:basic} we deliberately kept the discussion arbitrary, to demonstrate the robustness of the thermal instability. However, in order to run simulations that are applicable to real protoplanetary discs, we must be specific. The CO snow-line is observed to occur in the outer regions of protoplanetary discs \citep{Qi_science,Guidi2016,Qi2019}, thus we initially choose to explore the CO snow-line in this work. 

\subsection{Disc temperature calculation}
Our mechanism requires that we dynamically evolve the temperature structure of the disc, along with the surface densities and ice/vapour compositions. Further, our discussion in Section~\ref{sec:basic} requires this calculation takes into account the non-grey nature of the radiative transfer. The typically used hybrid-grey FLD method \citep[e.g.][]{Kuiper2010,Bitsch2013} or other common 1+1D methods \citep[e.g.][]{dalessio98,dalessio01}, do not capture the physics of this thermal instability. The coupling of the thermal evolution of the disc to its dynamical evolution requires a simplified radiative transfer scheme. Therefore, we adopt the two-layer model \citep[e.g.][]{chiang97,Chiang2001,Dullemond2001}, where the disc is separated into a ``super-heated'' atmosphere (heated by direct irradiation from the star) at a temperature of $T_{\rm atm}$ and an interior (heated by the reprocessed stellar irradiation emitted by the disc's atmosphere) at a temperature of $T_{\rm m}$. Within the interior and atmosphere the temperatures are taken to be constant and independent of height. The interior is taken to be in vertical hydrostatic equilibrium with this temperature profile. The transition from the interior to the atmosphere occurs at the photosphere to stellar irradiation ($h$).\footnote{Note in this work we use the opposite, but more common, convention to \citet{Chiang2001}, where we use $H$ for the disc's scale height and $h$ for the height of the photosphere.}  This scheme has been shown to accurately capture the nature of the radiative transfer problem when compared to more complete radiative transfer solutions \citep[e.g.][]{Dullemond2003,Rafikov2006}. 

The two-layer scheme as proposed by \citet{chiang97}, was later modified by \citet{Chiang2001} and \citet{Dullemond2001}. We build on these previous works and present a slightly different formalisation, suitable for coupling within a dynamical calculation. To avoid slow integration over frequency (e.g. Equations~\ref{eqn:heat} \& \ref{eqn:cool}) we follow \citep{Chiang2001} and consider three average radiation fields and mean opacities. Firstly, the opacity of the atmosphere to stellar light is described by a Planck-mean opacity at the stellar temperature ($\kappa_p(T_*)$). Secondly, the opacity to radiation at the disc's atmosphere's temperature is described by a Planck-mean opacity at the temperature of the atmosphere ($\kappa_p(T_{\rm atm})$). Finally, we approximate the opacity to radiation at the interior's temperature by a Planck-mean opacity at $T_m$. 

These simplifications allow the temperature of the atmosphere to be written as:
\begin{equation}
    T_{\rm atm} = T_* \left(\frac{R_*}{R}\right)^{1/2}\left( \frac{\phi\kappa_p(T_*)}{4 \kappa_p(T_{\rm atm})}\right)^{1/4}
\end{equation}
where $R_*$ is the stellar radius, $R$ is the cylindrical radius from the star and $\phi$ is the fraction of the stellar disc seen by the atmosphere which we set to $1/2$. The heating rate per unit area (including both sides of the disc) is given by:
\begin{equation}
    Q_+ = \phi \left[1-\exp\left(-\frac{\Sigma_d \kappa_p(T_{\rm atm})}{2}\right)\right]\left(\frac{R_*}{R}\right)^2\sigma_b T_*^4 \sin \alpha_{\rm f} \label{eqn:heat_use}
\end{equation}
where $\alpha_f$ is the ``flaring'' angle - i.e. the grazing angle between incoming stellar irradiation and the photosphere. The factor of a half in the exponential comes from the fact the irradiation from the disc's surface only traverses half of the disc's surface density before it reaches the mid-plane.

Finally, the cooling-rate per unit area (including both sides of the disc) is:
\begin{equation}
    Q_- = 2  \left[1-\exp\left(-\frac{\Sigma_d \kappa_p(T_m)}{2}\right)\right] \sigma_b T_m^4  \label{eqn:cool_use}
\end{equation}
\subsubsection{Calculation of the flaring angle}
Calculation of the flaring angle is non-trivial. An implicit ray-tracing can result in numerical instabilities \citep[e.g.][]{Dullemond2000}. In the limit that $\alpha_f$ is small, the definition of the flaring angle \citep[e.g.][]{chiang97} becomes:
\begin{equation}
    \alpha_f = \left(\frac{3}{4\pi}\frac{R_*}{R}\right) + R \frac{\rm d}{{\rm d}R}\left(\frac{h}{R}\right) \label{eqn:alpha1}
\end{equation}
Locally, ignoring shadowing (see Section~\ref{sec:shadows} for a discussion of shadowing), the flaring angle is also defined by, \citep{Dullemond2001}:
\begin{equation}
    \frac{\kappa_p(T_*)}{\alpha_f}\int_h^\infty \rho(z) {\rm d}z = 1
\end{equation}
Assuming the vertical density structure is Gaussian (as true for an isothermal disc in hydrostatic equilibrium)  this can be solved to give, \citep{Dullemond2001}:
\begin{equation}
    1-{\rm erf}\left(\frac{h}{\sqrt{2}H}\right)=\frac{2\alpha_f}{\Sigma_d\kappa_p(T_*)} \label{eqn:alpha2}
\end{equation}
We note in passing that Equations~\ref{eqn:alpha1} \& \ref{eqn:alpha2} can be combined to create an ODE for $h$. However, in practice solving this ODE requires knowing the height of the photosphere at the inner boundary. We find that we cannot accurately estimate this boundary condition such that integration of the ODE provides a numerically stable solution for $\alpha_f$. Therefore, following \citet{Dullemond2001} we note that $h$ is a smoothly varying function and write that $R{\rm d}/{\rm d}R \approx \gamma -1$ in Equation~\ref{eqn:alpha1}, where $\gamma$ is the ``flaring index of the disc'' formally defined by:
\begin{equation}
    \gamma \equiv \frac{{\rm d}\log h}{{\rm d} \log a}
\end{equation}
Following both \citet{Chiang2001} and \citet{Dullemond2001}, we note that $h/H$ is an extremely slowly varying function (roughly varying from 5 to 3 from $<0.1$~AU to $>100$~AU); thus, we approximate $\gamma$ as:
\begin{equation}
    \gamma \approx \frac{{\rm d}\log H}{{\rm d} \log a}\label{eqn:gamma_approx}
\end{equation}
so the flaring angle becomes:
\begin{equation}
    \alpha_f \approx \left(\frac{3}{4\pi}\frac{R_*}{R}\right) +\left(\gamma -1\right)\frac{h}{R} \label{eqn:alpha_final}
\end{equation}
where the combination of Equation~\ref{eqn:alpha2} \& \ref{eqn:alpha_final} can be solved to find the value of $h$ and hence $\alpha_f$. We do this using Brent's root finding method with a relative tolerance of $10^{-6}$. This approximate form of the flaring index allows us to include both changes from the temperature profile and surface density. In the case that the flaring index does not change, this give us that the heating rate is $\propto T_m^{1/2}$ we used previously. 

%Previous works that have employed this method use the $\gamma$ value from the previous radial grid point and update it only every two radial grid point. We find this approach leads to large errors at opacity jumps (e.g. at snow-lines) and non-continuous temperature structures. Therefore, we evaluate $\gamma$ using 4th order numerical differencing of Equation~\ref{eqn:gamma_approx}; 2nd order differencing is used near the boundaries, and at the inner boundary the flat disc solution, $\gamma=1.125$ \citep{chiang2001} is used, and at the outer boundary $\gamma$ is simply chosen to be constant. When the temperature structure is solved for in an explicit or implicit time-dependant manner (as we shall do in Section~\ref{sec:Numerical}) this approach is numerically stable. Thus, our calculation of the flaring angle is (unlike previous ones) local in the sense that changes in surface density, opacity of temperature effect the flaring angle at the point. Additionally, unlike the method of \citet{chiang97} \& \citet{Chiang2001} we do not fix the height of the photosphere to be a fixed number of scale heights; hence the photospheric height can be effected by changes in the surface density locally (as occurs at snow-lines). However, as we shall discuss in Section~\ref{sec:discss} our method does not allow for shadowing, which as we shall see, should occur in these discs. 

\subsubsection{Latent Heat}

Any phase change carries with it an exchange of internal energy and thermal energy. The sublimation of ice requires the absorption of thermal energy, cooling the surroundings. Therefore the act of sublimation is thermally stabilising. Thus, the absorption/release of latent heat has the potential to stabilise our thermal instability. However, due to the short thermal timescales in the outer regions of protoplanetary discs this turns out not to be the case (at least for the CO snow-line). This result can be seen from the following analysis. 

The heating rate per unit area due to latent heat can be written as:
\begin{equation}
    Q_{\rm lat} = L_{\rm sub} \left(\frac{\dot{M}_{\rm CO}}{2\pi R}\right)\frac{{\rm d}X_{\rm ice}}{{\rm d}R}
\end{equation}
where $L_{\rm sub}$ is the latent heat of sublimation, $\dot{M}_{\rm CO}$ is the CO mass-flux and $X_{\rm ice}$ is the fraction of the volatile species in the ice phase. The radial derivative of the ice-mass fraction can be written as:
\begin{equation}
    \frac{{\rm d}X_{\rm ice}}{{\rm d}R} = \frac{{\rm d}X_{\rm ice}}{{\rm d}T_m}\frac{{\rm d}T_m}{{\rm d}R}
\end{equation}
where the derivative of the ice-mass fraction as a function of temperature is set by the dynamics of the phase change (see Section~\ref{sec:phase}). Measurements of the latent heat of CO for sublimation temperatures closer to 50-60~K indicate a value of $\sim 3\times10^9$~erg~g$^{-1}$ \citep{REF:NIST}. While the sublimation temperature under protoplanetary conditions is closer to 20-30~K, one would not expect the value of the latent heat to be dramatically different. 

We can estimate the role of cooling due to sublimation by considering its ratio with respect to radiative cooling (in the limit $\tau<1$):
\begin{eqnarray}
\frac{Q_{\rm lat}}{Q_{\rm cool}}&=& \frac{L_{\rm sub}X_{\rm ice}\dot{M}_{\rm CO}}{2\pi R^2 \tau \sigma_b T_m^4}\frac{\partial \log X_{\rm ice}}{\partial \log T_m}\\
&\approx& 2\times10^{-6} \,\,\,  \frac{X_{\rm ice}}{\tau} \frac{\partial \log X_{\rm ice}}{\partial \log T_m} \left(\frac{\dot{M}_{\rm CO}}{10^{-9}~{\rm M_\odot~yr^{-1}}}\right)\nonumber \\ &\times&\left(\frac{R}{50~{\rm AU}}\right)^{-2}\left(\frac{T_m}{25~{\rm K}}\right)^{-4}
\end{eqnarray}
Our CO sublimation model indicates an appropriate value of $\partial \log X_{\rm ice}\partial\log T_m \sim 10-30$ in the region of the snow-line. Therefore, even with our quite large choice of the CO mass flux (and the fact that the snow-line is only thermally unstable at intermediate optical depths) we find that latent heat is unimportant at the $\sim 10^{-5}$ level, and will not stabilise our thermal instability. This is born out when we explicitly include latent heat in the calculation of our CO S-curve and find it's unimportant. In our numerical simulations in Section~\ref{sec:Numerical} we do not include latent heat for simplicity. 

\subsubsection{Opacities}
The precise details of CO opacities has not been explored in as much detail as other ices (e.g. Water) or silicates. Therefore, we proceed to continue to work with opacties motivated from geometric optics. Firstly, for simplicity we consider an opacity law than encodes the minimum physics, namely we ignore the particle size dependence on the opacity and assume the Planck-mean opacity is given by:
\begin{equation}
    \kappa_p = 10~{\rm cm}^2~{\rm g}^{-1}\left(\frac{T}{10~K}\right) \label{eqn:simple_opac}
\end{equation}
We note that this is opacity {\it per unit mass of solids}, not per unit mass of gas (+ dust). Therefore, for a dust-to-gas mass ratio of 0.01 this opacity corresponds to the ``standard'' sub-mm opacity of $0.1$~cm$^{2}$~g$^{-1}$. 

We can explore the exact dependence on temperature and particle size by taking more considerations from geometric optics. Further, assuming an MRN grain-size distribution with the number density distribution scaling as $n(s){\rm d}s\propto s^{-3.5}{\rm d}s$, up to a maximum size of $s_{\rm max}$, the opacity is dominated by the smallest grains which we set to 0.1~$\mu$m. However, the mass is dominated by the largest grains. Such choices motivate a phenomenological opacity functional form of:
\begin{equation}
    \kappa_p(s_{\rm max},T) = \kappa_0\left(\frac{s_{\rm max}}{1~{\rm cm}}\right)^{-1} \frac{1}{\left(\mathcal{R}^{-\beta}+\mathcal{R}^{-1/2}\right)}\label{eqn:geometric_opacities}
\end{equation}
with $\mathcal{R}=\pi s_{\rm max} T/b$, roughly the ratio of the mean photon wavelength to the size of the particle ($b$ again is the constant in Wein's displacement law)\footnote{We find dropping the factor of 2 in this ratio gives a better fit to numerically calculated opacities.}. This form correctly reproduces the limits of the parameter space and the form of the opacity function. For $\beta=$ 1 \& 2, we find $\kappa_0 = 0.9~$cm$^{2}$~g$^{-1}$ \& $0.7~$cm$^{2}$~g$^{-1}$  produces a good match to numerically calculated mean opacities respectively. The choice of this functional form over tabulated opacities is purely one of numerical efficiency. Finally, for the choice of opacity in the atmosphere of the disc ($\kappa_p(T_{\rm atm})$) we assume only small particles are present (due to vertical settling) and set $s_{\rm max}=1~\mu$m in the disc's atmosphere. 

\subsection{CO ice-vapour chemistry}\label{sec:phase}
Our CO sublimation model follows \citet{Booth2017} and \citet{Booth2019}, which itself is based on the models of \citet{Hollenbach2009}. \citet{Booth2017} explicitly demonstrated the timescale for small solid particles (e.g. less than a few centimetres) to reach an adsorption-desoprtion equilibrium is short compared to the relevant transport timescale in a protoplanetary disc. This conclusion is in agreement with the results of \citet{Piso2015} who showed only particles larger $\sim 0.5$~cm drift appreciably while sublimating. Therefore, we solve for the equilibrium CO ice abundance by balancing the thermal adsorption and desorption rates in the mid-plane of the disc. The thermal desorption rate per grain is given by:
\begin{equation}
    \Gamma_d = 4\pi s_0^2 N_s R_d f_s
\end{equation}
where $N_s = 1.5\times10^{15}$~cm$^{-2}$ is the number of binding sites per unit area, $f_s$ is the fraction of binding sites covered by CO ice, $s_0$ is the size of the grain and $R_d$ is the desorption rate per molecule given by:
\begin{equation}
    R_d = \nu_i\exp\left(-\frac{T_{\rm bind}}{T_m}\right)
\end{equation}
with $\nu_i$ the vibrational frequency of the CO molecule in the potential well of the dust's surface given by, \citep{Hollenbach2009}:
\begin{equation}
    \nu_i = 1.6\times10^{11}~{\rm s}^{-1} \frac{T_{\rm bind}}{m_{\rm CO}/m_h}
\end{equation}
where $T_{\rm bind}$ is the binding temperature of the molecule to the surface in Kelvin. We use a value of $T_{\rm bind}$ = 960~K \citep{Aikawa1996}, which sets the sublimation temperature to $\sim 25$~K under typical disc conditions. 

Adsorption arises due to grain-molecule collisions; if all these collisions result in adsorption of the molecule, then the adsorption rate per grain is:
\begin{equation}
    \Gamma_a = n_{\rm CO,v}\pi s_0^2 v_t
\end{equation}
where $n_{\rm CO,v}$ is the number density of CO molecules in the vapour phase and $v_t$ is their thermal velocity. For the grain size we use $s_0=0.1~\mu$m; while in our work we do include a prescription for grain size we follow \citet{Booth2017}, who argue using the work of \citet{Stammler2017} that for a size distribution of $n(s){\rm d}s\propto s^{-3.5}{\rm d}s$ the adsorption-desorption equilibrium was set by the small grains (which dominate the total grain surface area), while the mass is dominated by the large grains. To find the CO ice and vapour fractions we simply equate the adsorption rate to the desorption rate.  

\subsection{Limit Cycle}\label{sec:CO_limit}
\begin{figure}
    \centering
    \includegraphics[width=\columnwidth]{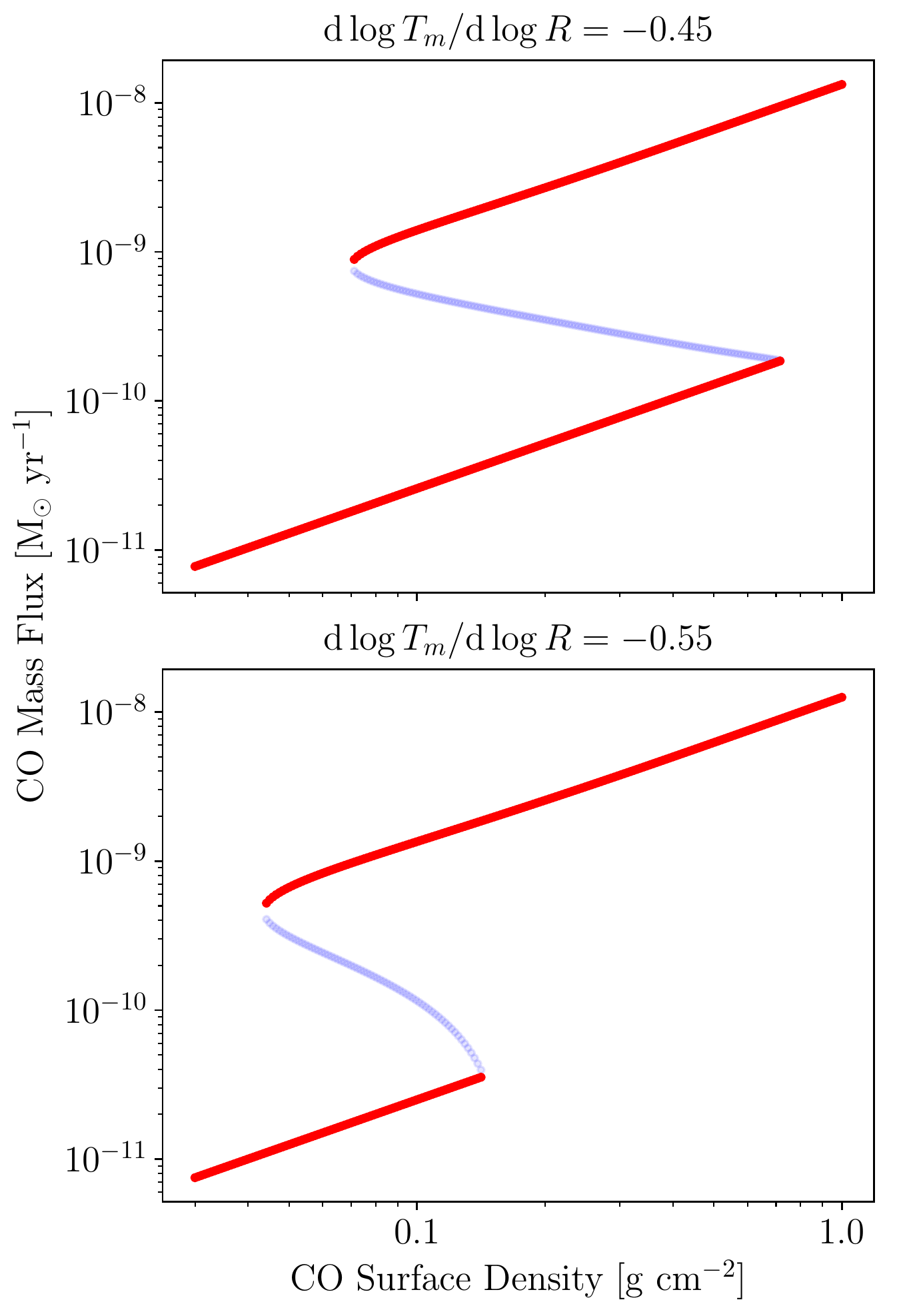}
    \caption{S-curves arising from the CO snow-line thermal instability. The dark-red line shows the thermally stable (vapour - low mass flux) and (ice - high mass flux) solutions. The light-blue line shows the thermally unstable solutions. Two panels demonstrate how sensitive the S-curve is to the temperature gradient in the disc. Steeper temperature gradients result in smaller changes in the CO surface density around the limit cycle. This S-curve shows there is clearly a large range of forbidden steady-state CO mass-fluxes (note the width of this forbidden region is entirely set by our choice of the ratio of the Stokes number to the viscous alpha).}
    \label{fig:CO_limit}
\end{figure}

Armed with the above methods for calculation of the disc's temperature and CO vapour and ice fractions we can calculate the real limit-cycle for the CO snow-line provided $\gamma$ is known. Here we simply pick a value for the temperature gradient (${\rm d}\log T_m / {\rm d}\log R =$ -0.45 and -0.55) and 
calculate the S-curve for the CO snow-line at 50~AU. We assume a solar-mass pre-main-sequence star with $T_*=4000~$K and $R_*=1.43~$R$_\odot$. For the ice-phase the particles are taken to have a stokes number of 0.05 (roughly 0.3 mm in size for a typical disc) and we assume a viscous $\alpha_v=10^{-3}$. The gas surface density profile is also taken to be $\Sigma \propto R^{-1}$, and the dust (excluding CO ice) surface density is taken to be 0.01 g~cm$^{-2}$. The above choices do not affect the existence of the limit-cycle and the thermal instability, but do affect the quantitative properties. Figure~\ref{fig:CO_limit} shows the CO S-curve. 
We have checked that the latent heat from CO sublimation does not change the S-curve at a bigger that the $10^{-5}-10^{-6}$  level. Figure~\ref{fig:CO_limit} bears a similarity to our simple example in Section~\ref{sec:basic} where there is a hot, vapour dominated, low mass-flux branch and a cold, ice dominated, high mass-flux branch. Unlike ``classical'' thermal instabilities, the size of the forbidden range of mass-flux is not set by the physics of the instability, rather the ability of the dust to drift relative to the gas. In a standard viscous accretion disc this is simply paramterised by the ratio of the Stokes number to the viscous alpha. The larger this ratio, the larger the forbidden region. 

However, the existence of a limit cycle and thermal instability is not fixed to the assumption of a viscous accretion disc. Rather it arises due to the non-grey radiative-transfer of how a protoplanetary disc is heated and the ability of the solids to drift relative to the gas. Therefore, in the case that protoplanetary accretion is wind-driven \citep[e.g.][]{Bai2013,Gressel2015,Bai2016}, provided the dust has grown to a sufficient size to drift relative to the gas, an S-curve and limit cycle will also exist.

\section{Numerical Calculations} \label{sec:Numerical}

Our discussion so far has focused on showing equilibrium steady-state solutions are unstable, and on the calculation of S-curves at fixed radii. Drawing on the analogy with previous thermal instabilities, we expect the disc to evolve through an ice-dominated ``high-mass-flux'' state and a vapour-dominated ``low-mass-flux'' state, where the snow-line propagates through the disc.  To calculate the disc's evolution and response to the thermal instability we perform numerical calculations where we couple the thermal calculation of the disc to the evolution of the solids and the ice-vapour partitioning of CO. To isolate different effects we fix the surface densities of gas, dust and CO as well as the maximum particle size at the outer boundary. This means, if there was no thermal instability, the simulation would then be expected to reach a steady state. We also work within the framework that the protoplanetary disc is a viscous accretion disc.

\subsection{Methods}

The numerical calculation for the evolution of the surface-densities follows the first-order in time, second-order in space algorithm presented in \citet{Owen2014b}. This is used to update the surface densities of CO-free dust, CO-ice and CO vapour through the standard thin-disc advection-diffusion equation for the surface density of species $\Sigma_i$:
\begin{equation}
    \frac{\partial\Sigma_i}{\partial t} = -\frac{1}{R}\frac{\partial}{\partial R} \left[R\Sigma_iv_i-\nu R \Sigma_g \frac{\partial}{\partial R}\left(\frac{\Sigma_i}{\Sigma_g}\right)\right]
\end{equation}
where $v_i$ is the velocity of the species, and $\nu$ is the kinematic viscosity\footnote{Hence we are implicitly setting the Schmidt number to unity.}. The velocity is set to the gas velocity for the CO-vapour. The dust-velocity for both the CO-free dust and CO-ice is (from the standard terminal velocity approximation in thin discs \citealt[e.g.][]{takeuchi02}):
\begin{equation}
    v_i = \frac{v_g\tau_s^{-1}-\eta \Omega R}{\tau_s+\tau_s^{-1}}
\end{equation}
with $\tau_s = \pi \rho_i s_{\rm max} / 2 \Sigma_g$ the Stokes number in the Epstien drag limit, and $\eta$ is a measure of the mid-plane gas pressure gradient given by:
\begin{equation}
    \eta = - \frac{{\rm d}\log P}{{\rm d}\log R}\left(\frac{H}{R}\right)^2
\end{equation}
\subsubsection{Gas Evolution}\label{sec:gas_evolve}
In our scenario where the dust and ice mass fraction is sufficiently small that the dynamical feedback on the gas is negligible, it is unclear whether the gas should evolve in the scenario where its surface density is fixed at the outer boundary. This depends on how you truly interpret the viscous alpha formalism. In the case $\nu = \alpha_v c_s H$, if $\alpha_v$ is a fundamental constant then the kinematic viscosity varies as the temperature changes and hence the gas-surface density will evolve. However, there is no reason to believe that $\alpha_v$ is {\it truly} constant. Therefore, we choose not to evolve the gas surface density and simply set $\nu = \nu_0 R$, as one would expect from a constant value of $\alpha_v$ if the temperature profile followed $T_m\propto R^{-1/2}$. For, ease of comparison $\nu_0$, is chosen such that if the disc had a temperature of 100~K at 1~AU and it obeys the $\alpha_v$ prescription, where $\alpha_v$ is the value we quote.

Therefore, in all our calculations the gas-surface density does not evolve and is fixed to follow:
\begin{equation}
    \Sigma_g = \Sigma_0 \left(\frac{R}{R_0}\right)^{-1}
\end{equation}

\subsection{Particle-size evolution}
It is well known that mutual collisions between dust particles results in growth in the outer regions of protoplanetary discs \citep[e.g.][]{Brauer2008,birnstiel10}. It is tempting to implement the well-used \citet{birnstiel12} dust-size evolution model. This model assumes that the sizes are set by a drift-growth equilibrium in the outer regions of a protoplanetary disc. However, this equilibrium is reached over length scales of $\sim R$, not length scales where the solid surface density can abruptly change at snow-lines. Implementation of this equilibrium model at the snow-lines results in rather abrupt particle size changes, which are unphysical. Therefore, motivated by the \citet{birnstiel12} model, we write the particle size evolution as:
\begin{equation}
    \frac{D s_{\rm max}}{Dt} = \frac{s_{\rm max}\Omega}{f_{\rm grow}}\frac{\Sigma_d}{\Sigma_g} \label{eqn:dust_size}
\end{equation}
where $f_{\rm grow}$ is a parameter that controls how efficient growth is (e.g. how efficiently particles stick when they collide); we choose $f_{\rm grow}=10$ \citep{Booth2020}. Equation~\ref{eqn:dust_size} reproduces the drift-growth equilibrium when the surface densities don't change abruptly, but doesn't introduce artificial changes in particle sizes at snow-lines. Equation~\ref{eqn:dust_size} is integrated using the same method as those for the surface densities: a scheme that is first-order in time and second-order in space. The advective term is integrated using a van-Leer limiter. We also implement a time-step criteria that the time-step must be shorter than 2\% of the growth-timescale.  

\subsection{Temperature and Ice-Vapour solver}
As discussed previously $\partial \log X_{\rm ice}/\partial \log T_m$ can be as large as $\sim 10-20$. Therefore, small changes in temperature can result in large changes in the ice-fraction and hence optical depth. Further, the thermal timescale is typically much shorter than the secular time-scales on which the dust grows and drifts. Therefore, the temperature evolution is typically solved implicitly and we do the ice-vapour balance inside this implicit temperature solution. This is done so the optical depth is correctly updated. Without this coupling inside the implicit temperature solver we find numerical instabilities in the ice-vapour fractions. 

Additionally, at the snow-line the radial component of the radiative flux may no longer be small compared to the vertical component (as typically assumed in thin disc theory). This effect arises due to large radial temperature or opacity changes. Such a similar effect is seen in other thermal instabilities and the radial component of the flux is explicitly included in numerical calculations \citep{Faulkner1983,Cannizzo1993}. This radial flux is negligible everywhere but at the snow-line. Thus, evolution of the mid-plane temperature follows, \citep[c.f.][]{Owen2014}:
\begin{equation}
    C_p \Sigma_g \frac{\partial T_m}{\partial t} = Q_{\rm heat}-Q_{\rm cool} - 2\left(\frac{H}{R}\right)\frac{\partial}{\partial R}\left(F_R\right)+Q_{\rm background}\label{eqn:thermal_evolve}
\end{equation}
where $F_R$ is the radial (cylindrical) radiative flux and $Q_{\rm background}$ is a background heating term from the surrounding star cluster. This background heating is set to a value of:
\begin{equation}
    Q_{\rm background} = 2 \left[1-\exp\left(\frac{\Sigma_d\kappa_p(10~{\rm K})}{2}\right)\right]\sigma_bT_{\rm 10\,{\rm K}}^4
\end{equation}
In the absence of other heat sources this heating term bathes the disc in a background radiation field of 10~K preventing the disc from becoming too cold in the outer regions.

As the instability we are interested in explicitly occurs at moderate optical depths, applying the radiative diffusion limit for $F_R$ is not correct. Limited with poor options about how to solve for $F_R$, we choose flux-limited diffusion. In flux-limited diffusion the flux is written as:
\begin{equation}
    F_R = - \frac{\lambda c}{\kappa \rho_m} \frac{\partial E}{\partial R}
\end{equation}
where $E_R=aT_m^4$ is the radiative energy density, $c$ is the speed of light, $\rho_m$ is the mid-plane density and $\lambda$ is the flux-limiter. The flux-limiter naturally blends the optically thick and thin region with a phenomenological smoothing function. Following \citet{Kuiper2010}, we choose:
\begin{equation}
    \lambda = \frac{2 + r }{6 + 3r+r^2}
\end{equation}
with $r=|\nabla E_R|/(\kappa \rho E_R)$, a measure of how optically thick the region is. We are now faced with a choice as to how to interpret $r$ within our thin disc model as it depends on $|\nabla E_R|$ rather than $\partial E_R/\partial R$. Ignoring this fact delivers the wrong approach to the limits. Through experimentation we find that the radial-flux is always sub-dominant (although not negligible) at snow-lines, therefore we crudely approximate $|\nabla E_R|$ as resulting from its gradient in the vertical direction only, so $\sim E_R/H$. This allows us to write $r\approx4/(\kappa(T_m)\Sigma)$. 

None of the above choices are thoroughly satisfactory; however, without a full 2D non-grey solution to the radiative transfer problem this is perhaps the most sensible method. The equations for heating and cooling rates are taken from Equations~\ref{eqn:heat_use} and \ref{eqn:cool_use} respectively. The heating rate depends on $\gamma$ which in turn depends on the temperature gradient (Equation~\ref{eqn:gamma_approx}) and the radiative flux term also includes terms that depend on the temperature gradient. Thus, Equation~\ref{eqn:thermal_evolve} cannot be solved at each annulus independently. Previous works that have employed this method use the $\gamma$ value from the previous radial grid point, and update it only every two radial grid points \citep{Chiang2001}. We find this approach leads to large errors at opacity jumps (e.g. at snow-lines) and non-continuous temperature structures. Therefore, we evaluate $\gamma$ using 4th order central numerical differencing of Equation~\ref{eqn:gamma_approx}, (2nd order differencing is used across the boundaries). When the temperature structure is solved for in an implicit time-dependant manner (as we do) this approach is numerically stable\footnote{We note the inclusion of the radiative diffusion term acts to also numerically stabilise the scheme.}. Thus, our calculation of the flaring angle is local in the sense that changes in surface density, opacity or temperature affect the flaring angle at that position. However, as we shall discuss in Section~\ref{sec:shadows} our method does not allow for shadowing, which as we shall see, should occur in these discs. 

Equation~\ref{eqn:thermal_evolve} is solved implicitly using the {\sc lsoda} numerical library for ODEs found in {\sc odepack} \citep{Hindmarsh1983}. We choose a relative tolerance for the implicit solver of $10^{-5}$ which provides a good balance between speed and accuracy. 

\subsection{Boundary conditions and numerical grid}
We are only interested in the physics of the snow-lines in the outer regions of the protoplanetary discs. However, truncating the grid inward of several AU results in an issue. In order to calculate the temperature outward of the inner boundary we need to know the temperature (or $\gamma$) inside the inner boundary. Simply guessing this temperature results in unphysical solution. Instead, if we pick our inner boundary to be on sub-AU scales, then the disc is in the flat-disc heating approximation, where the finite size of the star dominates the flux absorbed by the disc (the first term in Equation~\ref{eqn:alpha1}). In this limit $\gamma$ is known analytically to be 1.125 \citep{chiang97}. Therefore we chose our inner boundary to be at 0.04~AU. Setting $\gamma=1.125$ at the boundary allows us to solve for the temperature structure of the entire disc. Finally, $\gamma$ is chosen to be constant across the outer boundary. 

The large dynamical range in scales necessitates a logarithmic grid. We use 550 cells with an outer boundary of 300~AU. At the inner boundaries we use an outflow boundary condition for all the species and the radial radiative flux is set to zero.  At the outer boundaries the particle size is set to be in drift-growth equilibrium for a $T_m\propto R^{-1/2}$ temperature profile. For the radial radiative diffusion term, $H/R$ is assumed to be constant across the boundary. 

Our requirement to include the inner disc for the temperature calculation causes in two issues that produce small time-steps. Firstly the small grid cells and large particle sizes that can be reached in the inner disc due to continued growth (as we have not included fragmentation in our size evolution model), produce rapid drift with short time-steps. To mitigate this, once the dust size has grown to 5~cm it cannot grow further (this happens inside a few AU where the disc is optically thick and therefore insensitive to the actual particle size). Further, the small cell sizes in the inner disc result in short viscous timescales and consequently short time-steps. To mitigate this, inside an AU we smoothly reduce the viscosity by two orders of magnitude. Again as the disc is optically thick this has no effect on our calculations, and we note again the gas profile is fixed and does not evolve. With these choices the time-step allows manageable computations over Myr timescales, but still includes the full disc for the radiative transfer calculation. 

\subsection{Results}\label{sec:results}
As expected at intermediate optical depths, we find that the CO snow-line is thermally unstable and evolves through a limit-cycle where parts of the disc periodically condense and sublimate. Perhaps one of the most extraordinary findings is the scale over which this can happen. We find the CO snow-line can move 10s of AU! We find this behaviour to be robust to choices of opacity-laws, viscosity, disc properties and numerical parameters. We run all our simulations for several ~Myr and find that they do not relax to a steady-state, rather they enter a regime where they repeatedly evolve through a limit-cycle. 

\begin{figure*}
    \centering
    \includegraphics[width=\textwidth]{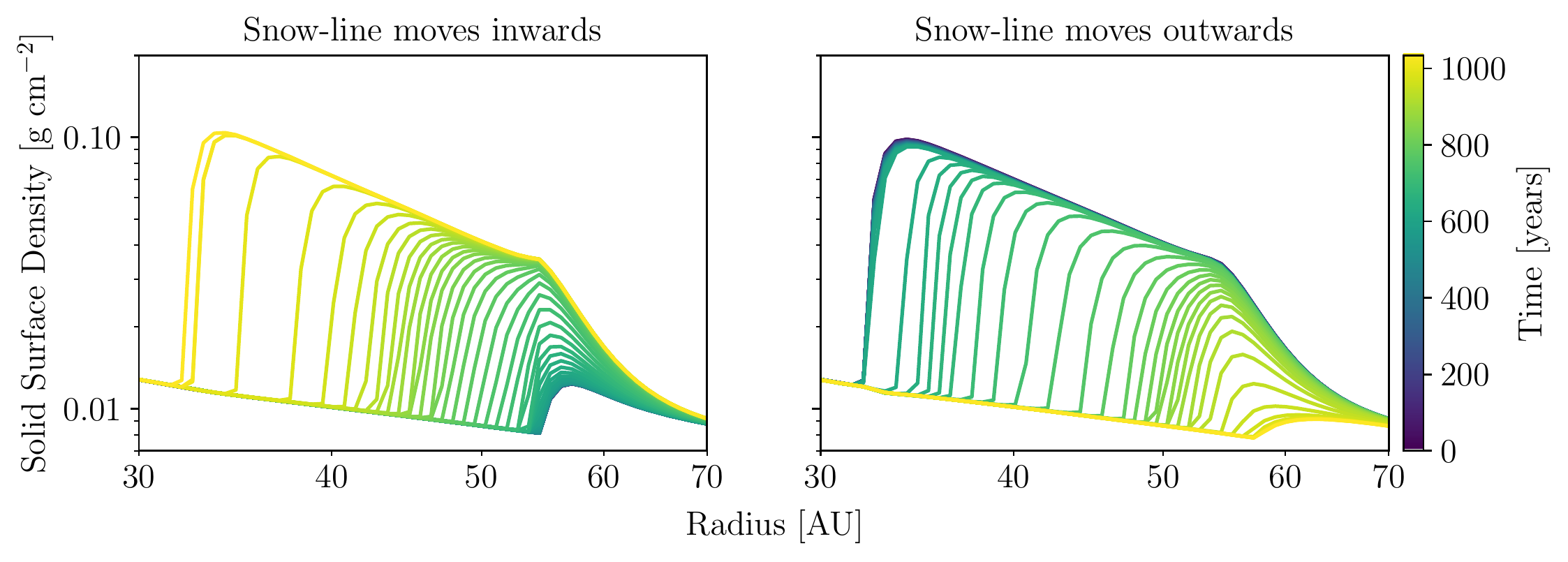}
    \caption{Snapshots of the solid (CO ice + non CO ice) surface density as a function of time. The left panel shows the snow-line advancing inwards. The right panel shows the snow-line retreating outwards.}
    \label{fig:burst_evolve}
\end{figure*}

%\subsubsection{Simple Opacity Laws}
Working with our simple opacity law (Equation~\ref{eqn:simple_opac}), we consider our standard model to be a 0.035~M$_\odot$ gas disc where the outer regions have been depleted slightly, so that the dust-to-gas ratio at 300~AU is $1/300$ (note the dust-to-gas ratio typically increases inwards). We choose a viscous alpha of $10^{-3}$ and assume that at the outer boundary 20\% of the total solid's mass is in CO ice (disc evolution calculations based on static chemistry put this value in the range 20-50\%, e.g. \citealt{Madhusudhan2014}).  In Figure~\ref{fig:burst_evolve}, we show snapshots of the solid surface density as a function of time. 
This figure indicates that the snow-line advances from $\sim 55$ to $\sim 30$~AU over a timescale of order 1000~years. It then spends several thousands to 10s of thousands of years with the snow-line at $\sim$ 30~AU before retreating back to $\sim 55$~AU on a timescale of 1000~years. During this time the solid-surface density in the region of 30-50~AU increases by an order of magnitude.  

\begin{figure*}
    \centering
    \includegraphics[width=\textwidth]{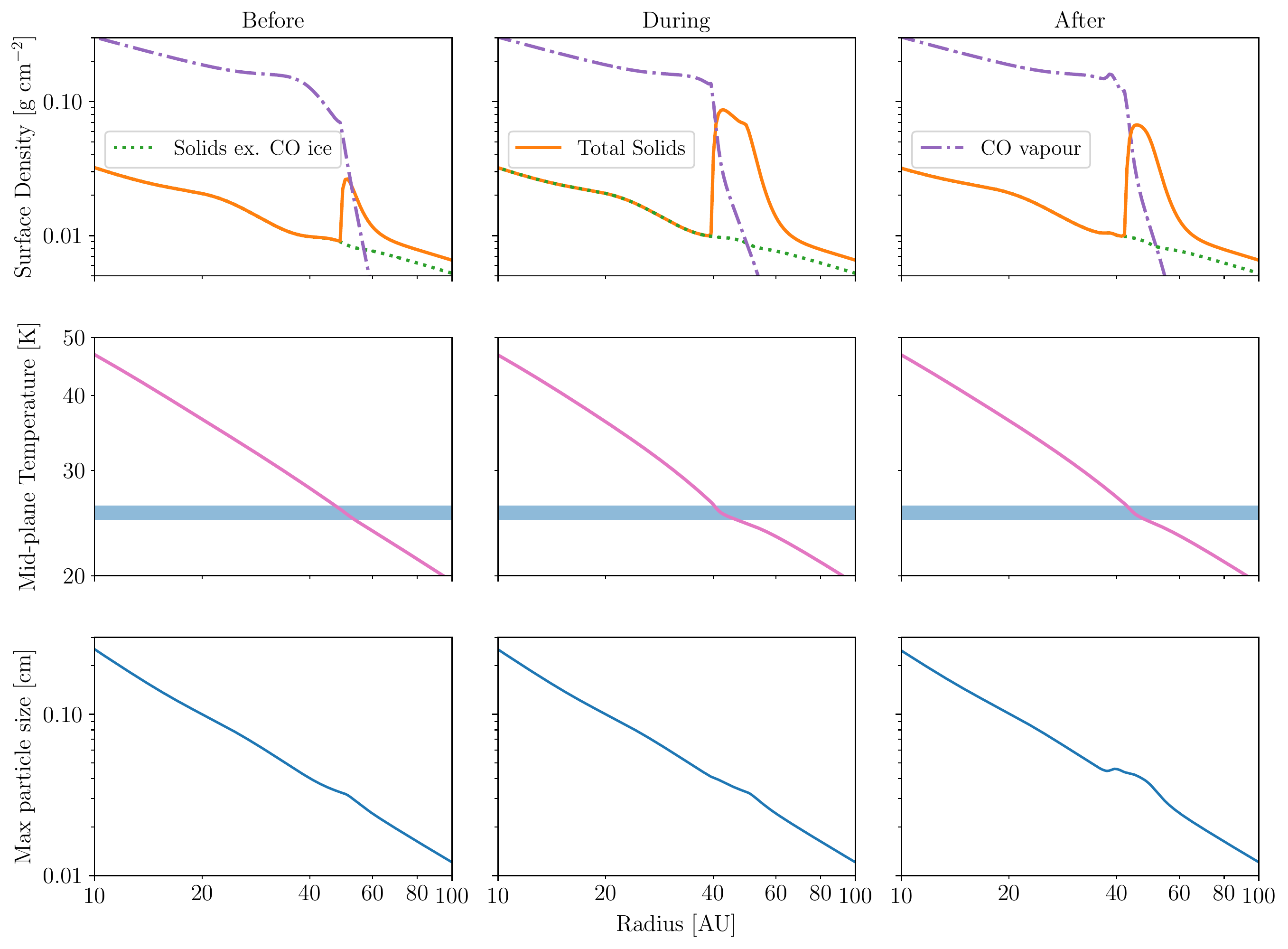}
    \caption{The top row shows the solid surface density, solid surface density excluding CO ice and CO vapour surface density. The middle and bottom rows show the mid-plane temperature and maximum particle sizes respectively. The left column shows the disc before the snow-line starts to rapidly move inwards, the middle shows the snow-line while it is moving inwards rapidly and the right shows while it is moving outwards. The horizontal band in the temperature panel shows the range of temperatures over which the ice fraction changes significantly and is representative of the CO sublimation temperature. }
    \label{fig:panel}
\end{figure*}

This rapid increase in solid surface density is easy to understand by looking at Figure~\ref{fig:panel} where we plot the temperature and maximum particle size as well as the solid surface density, solid surface density excluding CO ice and CO vapour surface density. The three columns show these quantities before the snow-line rapidly moving inwards (left), while it is moving inwards (middle) and while it is moving outwards (right). 
The CO vapour density rapidly increases inside the snow-line. This is well known \citep[e.g.][]{Stevenson1988}, as the dust drifts rapidly with respect to the gas, once the CO sublimates it moves much more slowly at the gas velocity. Thus, mass-conservation requires a much higher surface density. Once the snow-line moves in it progressively condenses the CO vapour resulting in high solid densities. 

One of the intriguing consequences is the solid surface density changes inwards of the snow-line (around 20~AU) in the top panels of Figure~\ref{fig:panel}. The consequence of these perturbations can be understood by looking at the evolution of the mid-plane temperature. As the snow-line moves inwards it flattens the temperature profile in the vicinity of the snow-line (middle panel of Figure~\ref{fig:panel}). A flatter temperature profile creates a weaker mid-plane pressure gradient and slow solid drift velocities. Thus, CO-ice free solids pile-up at the snow-line creating a density perturbation that then drifts inwards once the snow-line recedes and the drift speed increases. Dust diffusion smears out the density bumps as they drift inwards, but they can still be seen inwards of 10~AU, with, in some cases, order unity changes in the dust surface density. Therefore, not only do you get solid surface-density perturbations at the snow-line, but also perturbations at smaller radii that were created by previous movement of the snow-line. We find this effect is more pronounced at higher dust-to-gas ratios and CO-ice fractions (see Figure~\ref{fig:down-stream-bumps}). 

Finally, the bottom row of Figure~\ref{fig:panel} also shows the maximum particle size is only marginally affected by the change in solid surface density and movement of the snow-line. This is primarily because the change in surface density and movement of the snow-line is comparable or shorter in timescale to the growth timescale. 

\begin{figure}
    \centering
    \includegraphics[width=\columnwidth]{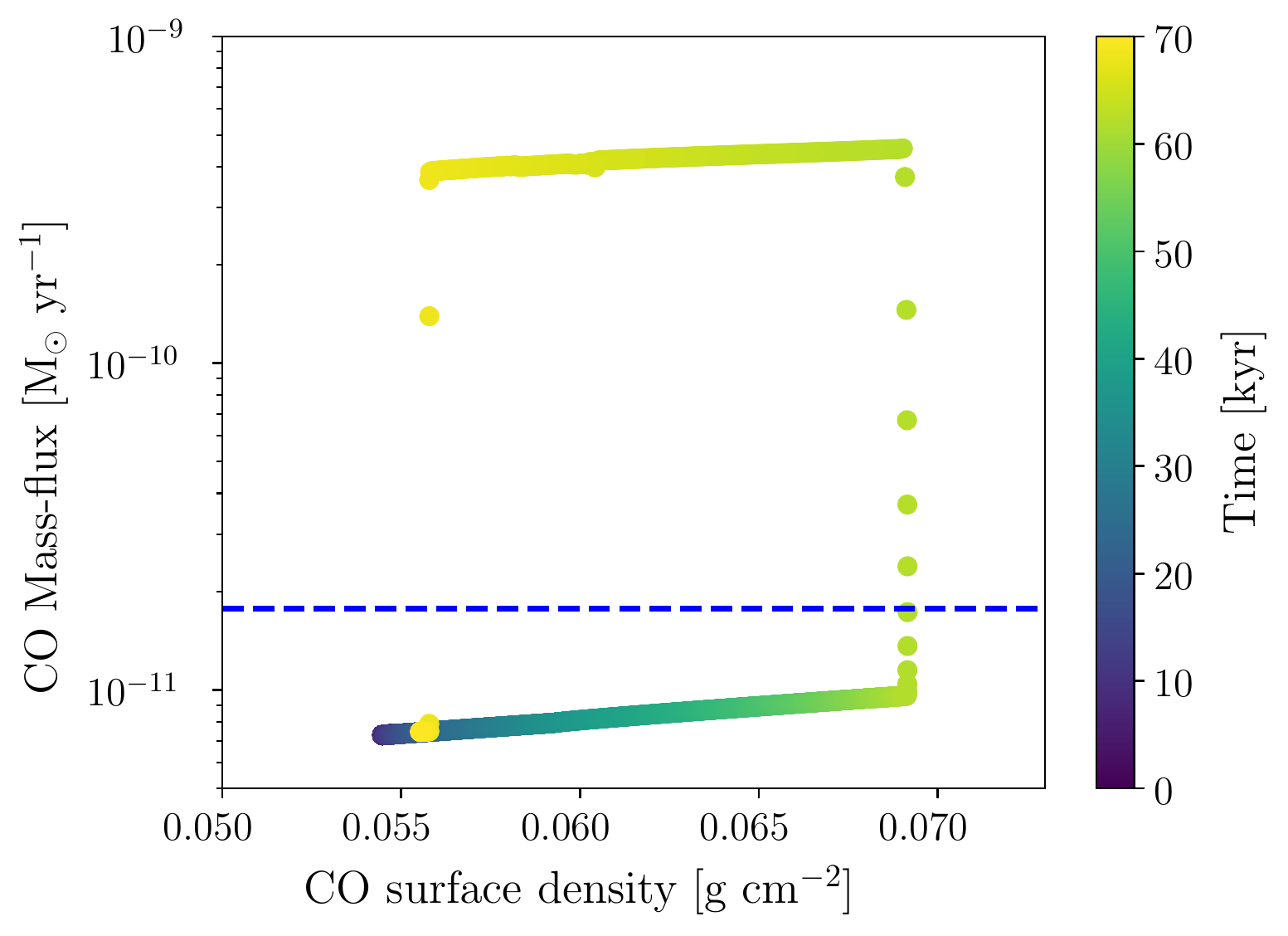}
    \caption{The evolution of the disc at 50~AU in the CO surface density -- mass-flux plane. The line traces out the limit cycle performed by this patch of the disc over one cycle. The overall cycle takes approximately 70,000 years. The colour of the points map the time evolution, where the bottom-left of the cycle is defined to have zero time. The dashed line indicates the CO mass flux at which the disc is being fed at the outer boundary.}
    \label{fig:actual_limit}
\end{figure}

\begin{figure}
    \centering
    \includegraphics[width=\columnwidth]{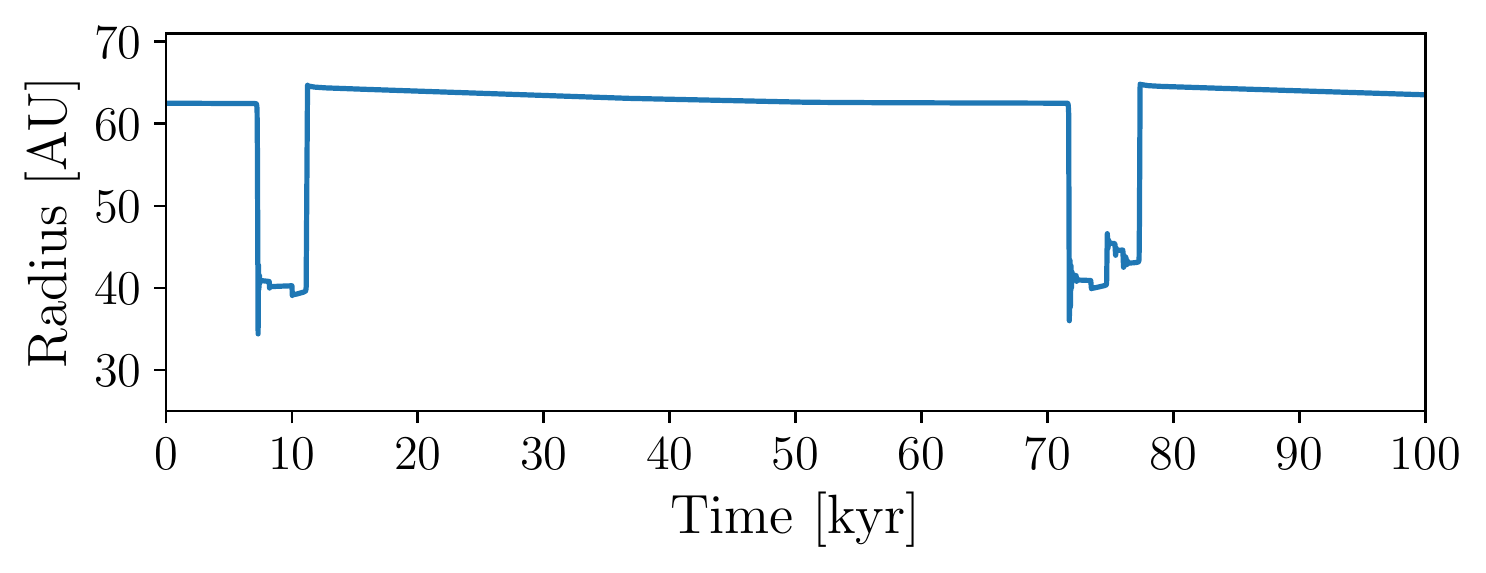}
    \caption{The evolution of the snow-line (taken to be the location where $X_{\rm ice}=0.5$) position as a function of time. While individual snow-line evolutions are all slightly different the overall pattern is regularly repeated, with a cycle length of $\sim$ 70 kyr.}
    \label{fig:radius_evolve}
\end{figure}

We can investigate how well our simple picture developed previously works in practice by plotting how our numerical model traverses the CO surface density and CO mass-flux plane. In Figure~\ref{fig:actual_limit} we plot the actual limit-cycle the disc performs at 50~AU. 
The limit cycle looks qualitatively similar to those shown in Figure~\ref{fig:CO_limit}. The CO mass-flux at which the disc is being fed lies in the forbidden region of the S-curve. The evolution around the limit-cycle indicates that the 50~AU annulus spends the majority of it's time ($\sim50~$kyr), in the vapour phase and the rest in the solid phase. Therefore, the duty cycle for this set of parameters, is such that the snow-line spends the largest fraction of it's time at larger radii, before propagating inwards and spending a smaller fraction of its time at closer orbital distances. This is shown in Figure~\ref{fig:radius_evolve} where we plot the snow-line position as a function of time.  

The narrow range of CO surface densities which the limit-cycle traverses is indicative of the steepness of the mid-plane temperature profile (${\rm d}\log T_m/{\rm d}\log R \sim -0.5-0.6$), where steeper temperature gradients give rise to smaller dynamic ranges in the limit cycle (Figure~\ref{fig:CO_limit}). We cannot compare our numerical evolutionary calculations to any steady-state S-curve, as the background temperature profile evolves throughout the evolution of the disc (middle-panels of Figure~\ref{fig:panel}). Further, we note that that the disc does not perform a perfect limit-cycle. The bottom-left of Figure~\ref{fig:actual_limit} indicates that once the CO sublimates into the CO vapour phase (the drop from high CO mass fluxes to low) the disc then evolves to lower CO surface densities, before then evolving to high CO surface densities as expected. This behaviour is generic, and is driven by diffusion. Once the CO locally sublimates into the vapour phase there is an over-concentration of CO vapour which diffuses away, lowering the CO surface density. 

\subsubsection{Small parameter exploration}
The parameter space to explore for this new thermal instability is vast, even if we restricts ourselves to just the CO snow-line. Here we only explore the dependence of several key parameters. 

Firstly, our understanding of the thermal instability is that if we move to low or high optical depths then the snow-line would indeed be thermally stable. We can modify the optical depth at the snow-line by changing the dust-to-gas ratio at the outer boundary. In the case of our nominal model the optical depth while CO is in the vapour phase is $\sim 0.1$ and $\sim 0.8$ when CO is in the solid phase. Therefore, changing the dust-to-gas ratio at the outer boundary to values of $1/2000$ and $1/30$ we find stable, time-invariant snow-lines occurring around 50-60~AU.

\begin{figure}
\begin{tikzpicture}
  \node (img1)  {\includegraphics[width=0.927\columnwidth]{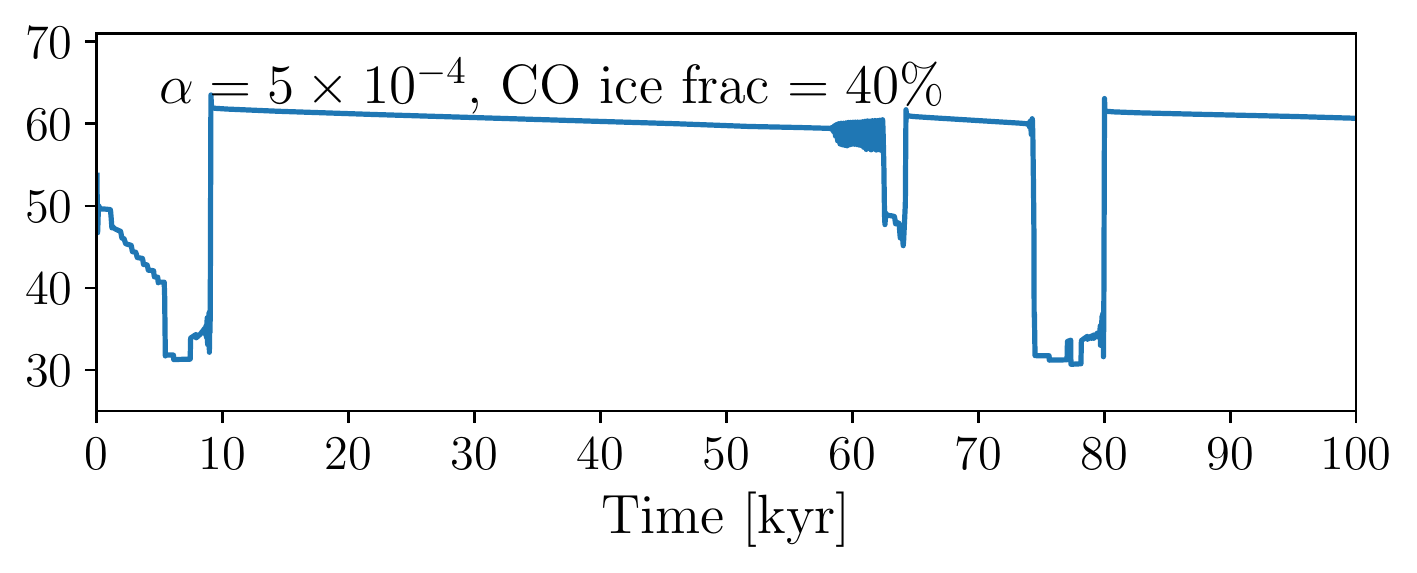}};
  %\node[below=of img1, node distance=0cm, yshift=1cm,font=\color{red}] {x-axis};
  %\node[left=of img1, node distance=0cm, rotate=90, anchor=center,yshift=-0.7cm,font=\color{red}] {y-axis};
  \node[above of = img1,yshift=1.8cm,xshift=-0.1cm] (img2)  {\includegraphics[width=0.9\columnwidth]{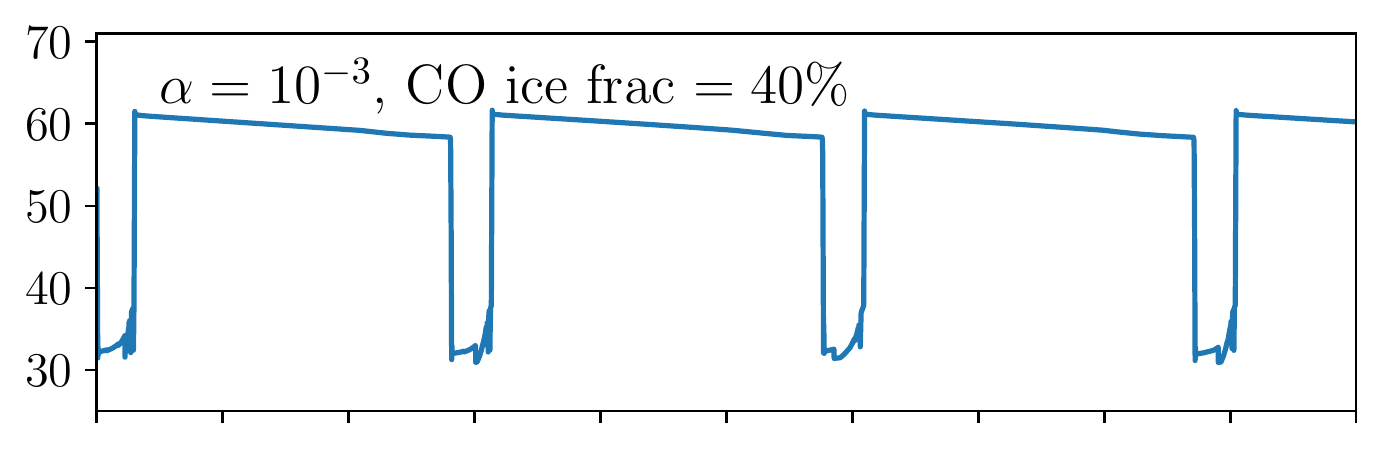}};
  \node[left of = img2, rotate=90, anchor=center,yshift=3.1cm,xshift=1.3cm]{Radius [AU]};
  \node[above of = img2,yshift=1.5cm] (img3)  {\includegraphics[width=0.9\columnwidth]{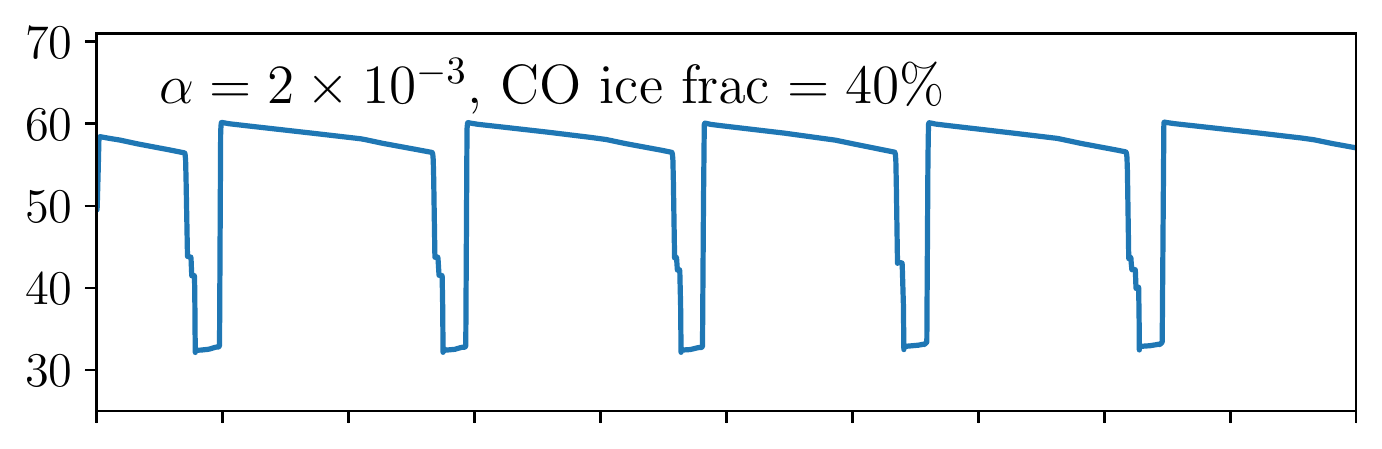}};
  \node[above of = img3,yshift=1.5cm] (img4)  {\includegraphics[width=0.9\columnwidth]{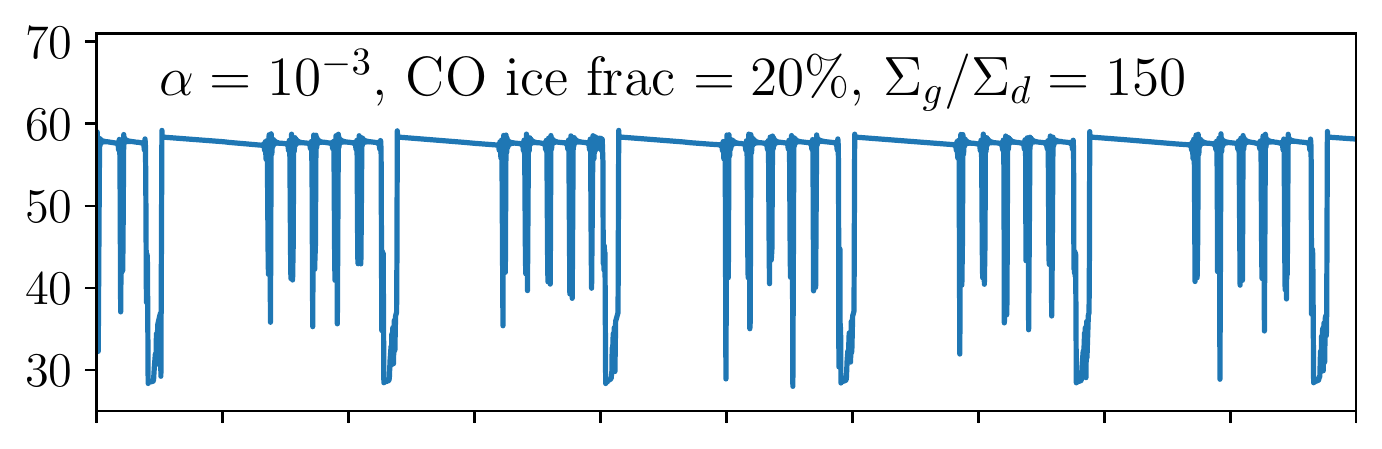}};
  %\node[below=of img2, node distance=0cm, yshift=1cm,font=\color{red}] {x-axis};
  %\node[left=of img2, node distance=0cm, rotate=90, anchor=center,yshift=-0.7cm,font=\color{red}] {y-axis};
\end{tikzpicture}
\vspace{-0.7cm}
\caption{The radius of the snow-line (taken to be the location where $X_{\rm ice}=0.5$) as a function of time, for different values of the disc parameters. This evolution is shown for a 0.1~Myr subset of the disc's evolution. While individual snow-line evolutions are all slightly different, the overall pattern is regularly repeated. The top panel is for a dust-to-gas mass ratio at the outer boundary of 1/150, while the bottom three panels are for the nominal choice of 1/300. The bottom three panels are for a CO ice fraction at the outer boundary of 40\%. These panels can be compared to Figure~\ref{fig:radius_evolve} for the nominal model with $\alpha=10^{-3}$, dust-to-gas mass ratio of 1/300 and CO ice mass fraction of 20\%.}
\label{fig:params_vary}
\end{figure}

Next we consider how changing the dust-to-gas ratio, the viscous alpha and the CO ice fraction controls the snow-line behaviour. The evolution of the snow-line radius as a function of time is shown in Figure~\ref{fig:params_vary} where the dust-to-gas ratio, viscosity and CO ice fraction are modified. The evolution of these different parameters can be compared to our nominal model shown in Figure~\ref{fig:radius_evolve}.

Due to the steep concentration gradients that occur at the snow-line, the primary driver, and the easiest way to understand the effect of changing disc conditions, is  diffusive transport across the snow-line. Larger concentration gradients, or higher values of the viscosity result in higher radial diffusive fluxes across the snow-line and faster evolution. Therefore, higher viscosities result in faster evolution around the limit-cycle while its properties are generally preserved (bottom three panels of Figure~\ref{fig:params_vary}). The higher CO mass-fraction between the third panel in Figure~\ref{fig:params_vary} and the nominal case (Figure~\ref{fig:burst_evolve}) similarly results in faster evolution around the limit-cycle due to the higher concentration gradients formed. 

\begin{figure}
    \centering
    \includegraphics[width=\columnwidth]{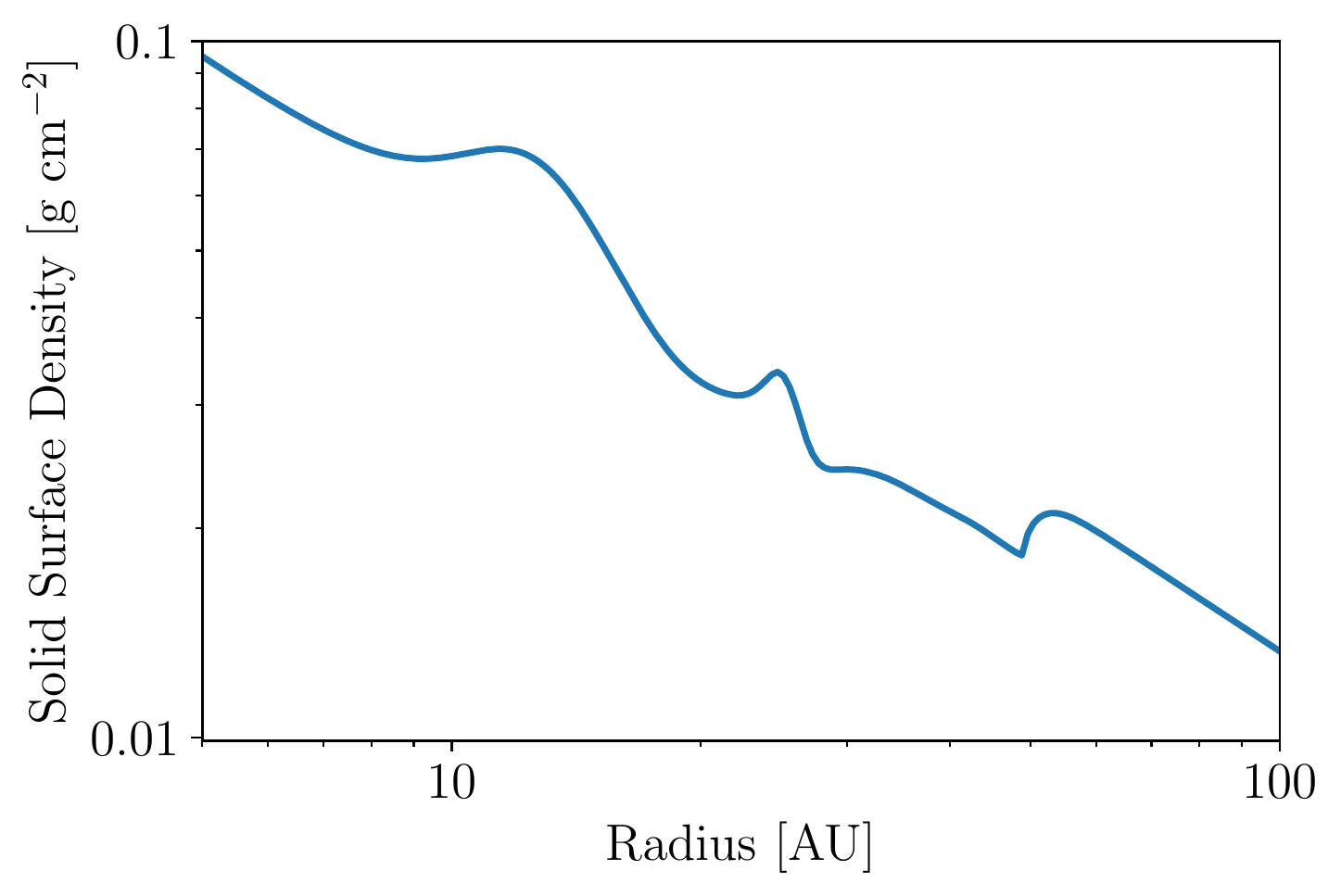}
    \caption{A snapshot of the solid surface density as a function of radius for the model with a dust-to-gas mass ratio of 1/150 at the outer boundary. The CO snow-line is at about 50~AU. This shows the formation of surface density perturbations interior to the snow-line as discussed in Section~\ref{sec:results}. Here we see two downstream surface density perturbations that originated at the CO snow-line when it moved inwards. The distance between bumps is set by how far the dust drifts after every repetition of the limit-cycle.}
    \label{fig:down-stream-bumps}
\end{figure}

Finally, the changes due to the dust-to-gas ratio can also be understood mainly in terms of radial transport across the snow-line (factors like the thermal timescale play a minor role). Higher dust-to-gas ratios result in faster particle growth. Larger particles drift faster with respect to the gas. Therefore, as the mass-flux is higher on the solid branch the evolution around the limit cycle is more rapid. Further, as the drift-speed of the particles with respect to the vapour in the gas is higher, a larger CO vapour gradient is formed across the snow-line. This results in higher diffusive transport and faster evolution around the limit cycle. We note that the higher dust-to-gas ratio model (and higher ice-fraction models) tend to produce more, and more pronounced, surface density perturbations interior to the snow-line which then drift towards the star (as discussed above). We show one such example in Figure~\ref{fig:down-stream-bumps} from the model with a dust-to-gas mass ratio of 1/150 at the outer boundary. This will obviously have implications for shadows which we discuss later in Section~\ref{sec:discuss}. 

\subsubsection{More physically motivated opacity laws}
\begin{figure}
    \centering
    \includegraphics[width=\columnwidth]{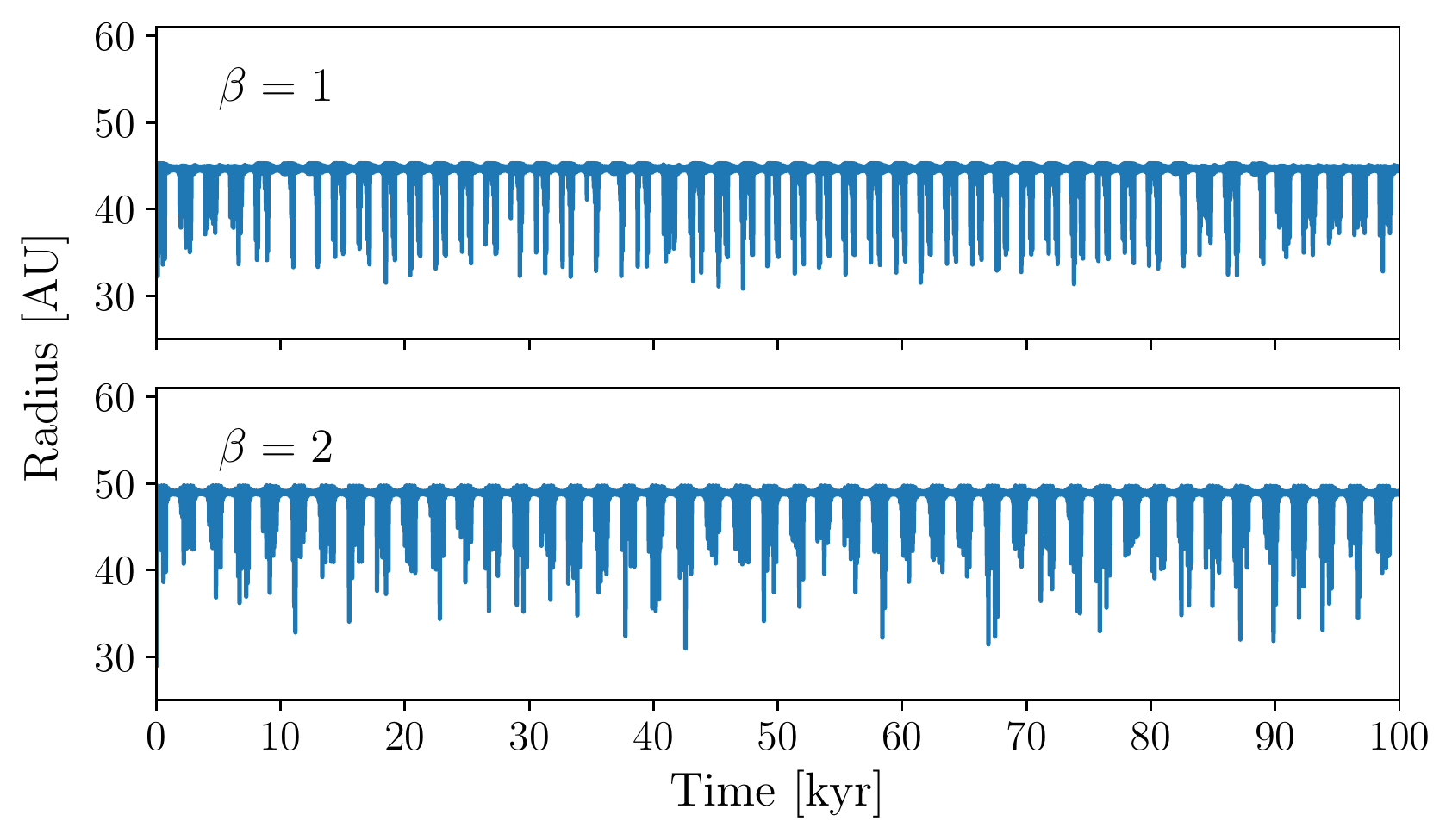}
    \caption{The evolution of the snow-line radius as a function of time for different values of the Rayleigh index $\beta$.}
    \label{fig:vary_opacity}
\end{figure}

In much of the previous analysis we have used a very simple opacity law that does not depend on the maximum particle size (Equation~\ref{eqn:simple_opac}). However, the opacity does vary with maximum particle size. Here we show results for opacity laws motivated from geometric optics (Equation~\ref{eqn:geometric_opacities}) where the Rayleigh index ($\beta$) is varied. Silicate grains have $\beta\sim 2$, whereas water-ice-silicate mixtures have lower values of $\beta$ \citep[e.g.][]{Chiang2001}. In Figure~\ref{fig:vary_opacity}, we show the snow-line evolution as a function of time for $\beta$ values of 1 and 2. 

This figure shows the evolution of the snow-line is faster and occurs over a smaller radial range than the cases with the simple opacity law. The changes are not expliclitly due to the differences in opacities themselves, but rather the effect they have on the dust-to-gas ratio and particle size. The modified opacities tend to result in steeper disc temperature gradients (hence a narrower radial range over which a snow-line can exist), and higher dust-to-gas ratios in the vicinity of the snow-line. Hence the snow-line evolution is faster, analogous to those models with higher dust-to-gas ratios with the nominal opacity (c.f. top panel Figure~\ref{fig:params_vary}).

\section{Discussion} \label{sec:discuss}

In this work we have shown that snow-lines in the outer regions (moderate optical depths) of protoplanetary discs are subject to thermal instabilities. The fact that the thermal instability only exists at moderate optical depths probably means that it will not affect the water snow-line under standard conditions. However, it is likely to control the dynamics of many other snow-lines that occur at cooler temperatures (e.g. carbon monoxide, carbon dioxide, ammonia etc). 

In the context of protoplanetary discs this thermal instability results in a limit-cycle where volatiles in the disc progressively sublimate and then condense. In keeping with other accretion disc thermal instabilities this limit-cycle can be analysed in the context of an ``S-curve'', where there is a forbidden range of volatile mass-fluxes. However, unlike other thermal instabilities the forbidden range of mass-fluxes is not set by the physics of the thermal instability itself. Rather, the existence of a limit-cycle only requires that dust can decouple, and drift relative to the gas, a requirement easily satisfied in the outer regions of protoplanetary discs. 

We have shown, even with a fairly small parameter study that the actual timescales and duty cycles for the evolution of the CO snow-line can vary significantly, with faster evolution occurring for higher viscosities, CO ice-mass fractions and dust-to-gas ratios. The simulations also indicate the duty cycle could be anywhere as low as $<10\%$ (as for our nominal model) or as high as $\sim 50\%$ (as in our models with more realistic opacities). Given we expect the dust-to-gas mass ratio (and possibly the CO ice-mass fraction) to vary as the disc evolves, we can expect the timescales associated with the limit-cycle to also evolve in time. We speculate the time to traverse the limit-cycle will increase as the disc evolves, as the dust-to-gas ratio falls in the outer disc with time. 

If these thermal instabilities were to operate in protoplanetary discs, the implications are far reaching. Progressive condensation of large amounts of volatiles as the snow-lines move inwards results in high dust-surface densities that may trigger planetesimal formation. The triggering of planetesimal formation could result in the sequestration of large amounts of volatiles (e.g. CO) in big, unobservable bodies, potentially providing a solution to the CO abundance problem in protoplanetary discs. This sequestration would then explain the mis-match between the HD and CO inferred gas masses \citep{Bergin2013,McClure2016,Powell2017,Bergin2018,Schwarz2018,Kama2020}. 

\subsection{Limitations}

We add several notes of caution to our work. This work has all been performed in a 1D model, where mid-plane values are taken to be representative of the disc. It is well known that snow-lines are two dimensional structures \citep[e.g.][]{Min2011,Qi2019}, and vertical mixing of both volatiles and solids are important \citep{Krijt2018}, and may qualitatively change our results. Since the majority of the mass is contained near the mid-planes of thin discs, we suspect our model does capture the basic physics, and it is unclear how 2D effects could stabilise a snow-line against this thermal instability. 

Further, we have implicitly assumed that the dust temperature is instantaneously transferred to the gas. In reality this change takes place over a finite time. This could potentially slow the evolution of our thermal instability. 

Finally, since we have assumed that the disc is being fed at a steady rate the rise in CO vapour surface density inside the snow-line extends all the way to the inner disc. Disc evolution models tend to find the CO vapour density is enhanced interior to the snow-line in a ring, rather than a continuous increase \citep[e.g.][]{Booth2017}. This may mean that the very high increase in ice mass, that occurs as the snow-line propagates inwards by 10s of AU, is overestimated.

\subsubsection{Are secular instabilities captured by secular models?}

Secular instabilities (i.e. instabilities that exist in the standard thin disc accretion models), such as our snow-line thermal instability, have a long history, and their true nature has been extensively debated \citep[as early as, e.g.][]{Pringle1981}. It is important to realise they are not ``real'' instabilities in the standard hydrodynamical sense. Therefore, it remains an open question as to how this thermal instability (and others) behave in real hydrodynamic discs. While there has been some success in recovering S-curve-like evolution in hydrodynamic models for other thermal instabilities \citep[e.g. dwarf nova outbursts,][]{Coleman2016}, these are not applicable to our snow-line thermal instability. For example, since we have temperature changes in our discs how do these affect the hydrodynamics of the underlying disc? Further, we note that the thermal timescales of the outer regions of protoplanetary discs can be fairly short, even comparable to the dynamical timescale ($\sim 1/\Omega_k$). Therefore, while we have assumed that the disc is in vertical hydrostatic equilibrium throughout our models (this is one of the basic assumptions of the thin disc model), this response may not be as instantaneous as we have assumed. Further, as discussed in section~\ref{sec:gas_evolve} we have not considered what happens if the viscosity changes due to the temperature changes, which would cause the gas disc to evolve. Nor have we included the dynamical feedback of the dust on the gas, which would also increase the richness of the problem. However, we end on a note of optimism: our calculations have shown that whatever happens, snow-lines are not stable structures and should evolve.

\subsection{Shadowing}\label{sec:shadows}
As discussed when we set-up our radiative transfer scheme, we are unable to include the effects of shadowing. Specifically, if the height of the photosphere blocks stellar light at more distant radii our model will not capture this. Since the evolving snow-line results in local surface density changes shadows are cast. As we are explicitly calculating the height of the photosphere, we can check whether our disc would have cast shadows. 

\begin{figure}
    \centering
    \includegraphics[width=\columnwidth]{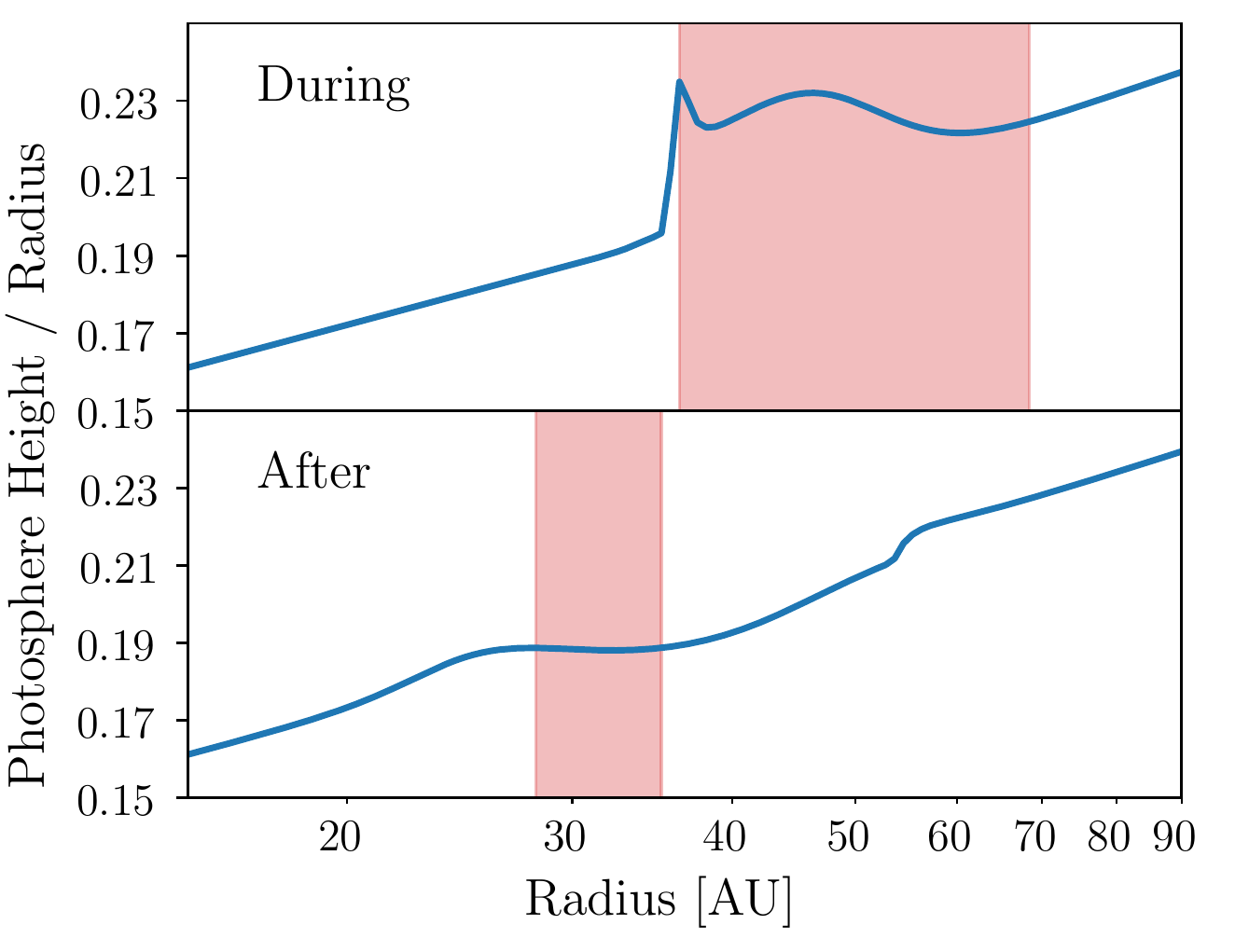}
    \caption{The ratio of the height of the photosphere to the cylindrical radius. If this ratio decreases with increasing radius the disc is shadowed (within the small angle approximation). The top panel shows the height of the photosphere a function of radius at a time when the snow-line has moved inwards and the bottom panel shows a case where the snow-line has just finished receding to large radius. The shaded regions show parts of the disc which will be shadowed. The top panel shows shadows cast by the snow-line itself and the bottom panel shows shadows cast by downstream dust surface density perturbations generated by the time variable snow-line.}
    \label{fig:shadows}
\end{figure}

We identify two types of shadow: firstly, shadows cast by the rapid increase in dust surface density exterior to the snow-line; secondly, the surface density perturbations that the moving snow-line creates in the down-stream dust surface density profile (Figure~\ref{fig:down-stream-bumps}) can also cast shadows. We show both these types of shadows in Figure~\ref{fig:shadows} computed from snapshots of our ``nominal'' case. 

Shadowing will result in a temperature minimum in the disc; similar to shadowing from the inner rim at the dust sublimation front \citep{Dullemond2001}, where the heating comes from the radial radiative flux and scattering. Therefore, in addition to the snow-line-created surface density structure, we speculate that these surface density structures will create further surface density structures due to pressure trapping and expulsion. Thus, there is potential that snow-lines can create many, time-variable, surface density structures. However, in order to make progress we either need to find a way of including shadows in our simple 1D radiative transfer scheme (something that has proved difficult in the past - e.g. \citealt{Dullemond2000}), or move to a 2D radiative transfer scheme, although it must be non-grey. 

\subsection{Observations of ringed discs}

Since the original high-resolution {\it ALMA} observation of HL Tau \citep{HLTau2015} there have been many observations of protoplanetary discs showing ringed sub-structure \citep[e.g.][]{Isella2016,Fedele2018,Andrews2018,Andrews2020}. While the community appears to favour a planetary origin for these structures \citep[e.g.][]{Zhang2018}, surface-density perturbations arising from snow-lines was among the ideas originally proposed \citep{Zhang2015,Okuzumi2016}. Recent work by \citet{Huang2018} and \citet{vanderMarel2019} has suggested that snow-lines cannot be the cause, since the observed structures do not appear to show preferred radial spacing ratios. Such preferred radial spacing ratios would be expected if discs had fixed radial temperature profiles (e.g. $T\propto R^{-1/2}$) and snow-lines occurred at a set temperature. 

We have shown snow-lines evolve in time and do not occur at fixed radial locations (especially in the outer regions of protoplanetary discs covered by the {\it ALMA} observations). Thus, such simple comparisons of radius ratios cannot be used to disprove or test the snow-line scenario. Further, even our simulations without shadowing show multiple, time-dependent structures from a single snow-line. The inclusion of shadowing is only likely to increase the number of ring-like structures arising from snow-lines. Therefore, while there is still much work to be performed, we speculate that the {\it ALMA} ringed discs could be caused by snow-line induced structures. Hence the observations are teaching us about chemistry and dust grain evolution in the outer regions of protoplanetary discs, rather than the properties of more mature planets. 

\section{Conclusions}
Snow-lines are the regions where volatile species rapidly transition from the solid phase to the vapour phase and typically occur in the outer regions of protoplanetary discs. The outer regions of protoplanetary discs are passivley heated by re-radiated stellar light from the disc's surface. The absorption of this re-radiated stellar light is primarily done by the solids. Since the temperature profile controls the position of the snow-line, and the position of the snow-line controls the temperature (through the distribution of solids), the radiative properties and chemistry of protoplanetary discs are inextricably linked. In this work, we have studied this coupling both analytically and numerically, indicating snow-lines can be thermally unstable. Our main conclusions are summarised below:
\begin{enumerate}
    \item Snow-lines are thermally unstable at intermediate optical depths ($\tau \sim 0.05 -2$). Therefore, many common snow-lines (CO, CO$_2$, NH$_3$, N$_2$) cannot exist at fixed radial locations and must evolve in time. As the water snow-line is at high optical depths ($\tau \gg 1$, and viscously heated) in standard protoplanetary discs it is likely to be thermally stable. 
    \item The unstable snow-lines evolve through a limit-cycle, where patches of the disc spend some of their time at cool temperatures, with the volatile species in the ice phase and some of their time at high temperatures, with the volatile species in the vapour phase. 
    \item This limit-cycle can be understood in terms of an S-curve, by an analogy to other accretion disc thermal instabilities. 
    \item Numerical models that explicitly couple the temperature structure to the distribution of solids using a non-grey radiative transfer scheme show the CO snow-line moves over 10s of AU on $\sim 1,000-10,000$~ year timescales. 
    \item The evolving snow-line creates ring-like structures interior to the snow-line that may be the origin of the observed {\it ALMA} structures. 
\end{enumerate}
While we have shown this snow-line thermal instability to be robust, the simplicity of our numerical model requires more work in the future. Priorities for future work are investigating the role of shadows cast by the time-evolving snow-line and investigating the 2D nature of the problem.

\section*{Acknowledgements}
I thank the referee for comments which improved the manuscript.
JEO is supported by a Royal Society University Research Fellowship. I'm grateful to Doug Lin for suggesting that the role of latent heat should be explored. I also thank Richard Booth, Ruth Murray-Clay and Giovanni Rosotti for useful discussions.

%%%%%%%%%%%%%%%%%%%%%%%%%%%%%%%%%%%%%%%%%%%%%%%%%%



%\bibliographystyle{mnras}

\bibliographystyle{mnras}
\bibliography{bib_paper}

\begin{thebibliography}{}
\makeatletter
\relax
\def\mn@urlcharsother{\let\do\@makeother \do\$\do\&\do\#\do\^\do\_\do\%\do\~}
\def\mn@doi{\begingroup\mn@urlcharsother \@ifnextchar [ {\mn@doi@}
  {\mn@doi@[]}}
\def\mn@doi@[#1]#2{\def\@tempa{#1}\ifx\@tempa\@empty \href
  {http://dx.doi.org/#2} {doi:#2}\else \href {http://dx.doi.org/#2} {#1}\fi
  \endgroup}
\def\mn@eprint#1#2{\mn@eprint@#1:#2::\@nil}
\def\mn@eprint@arXiv#1{\href {http://arxiv.org/abs/#1} {{\tt arXiv:#1}}}
\def\mn@eprint@dblp#1{\href {http://dblp.uni-trier.de/rec/bibtex/#1.xml}
  {dblp:#1}}
\def\mn@eprint@#1:#2:#3:#4\@nil{\def\@tempa {#1}\def\@tempb {#2}\def\@tempc
  {#3}\ifx \@tempc \@empty \let \@tempc \@tempb \let \@tempb \@tempa \fi \ifx
  \@tempb \@empty \def\@tempb {arXiv}\fi \@ifundefined
  {mn@eprint@\@tempb}{\@tempb:\@tempc}{\expandafter \expandafter \csname
  mn@eprint@\@tempb\endcsname \expandafter{\@tempc}}}

\bibitem[\protect\citeauthoryear{{ALMA Partnership} et~al.,}{{ALMA Partnership}
  et~al.}{2015}]{HLTau2015}
{ALMA Partnership} et~al., 2015, \mn@doi [\apjl] {10.1088/2041-8205/808/1/L3},
  \href {https://ui.adsabs.harvard.edu/abs/2015ApJ...808L...3A} {808, L3}

\bibitem[\protect\citeauthoryear{{Aikawa}, {Miyama}, {Nakano}  \&
  {Umebayashi}}{{Aikawa} et~al.}{1996}]{Aikawa1996}
{Aikawa} Y.,  {Miyama} S.~M.,  {Nakano} T.,   {Umebayashi} T.,  1996, \mn@doi
  [\apj] {10.1086/177644}, \href
  {https://ui.adsabs.harvard.edu/abs/1996ApJ...467..684A} {467, 684}

\bibitem[\protect\citeauthoryear{{Andrews}}{{Andrews}}{2020}]{Andrews2020}
{Andrews} S.~M.,  2020, arXiv e-prints, \href
  {https://ui.adsabs.harvard.edu/abs/2020arXiv200105007A} {p. arXiv:2001.05007}

\bibitem[\protect\citeauthoryear{{Andrews} et~al.,}{{Andrews}
  et~al.}{2018}]{Andrews2018}
{Andrews} S.~M.,  et~al., 2018, \mn@doi [\apjl] {10.3847/2041-8213/aaf741},
  \href {https://ui.adsabs.harvard.edu/abs/2018ApJ...869L..41A} {869, L41}

\bibitem[\protect\citeauthoryear{{Armitage}, {Livio}  \& {Pringle}}{{Armitage}
  et~al.}{2001}]{Armitage2001}
{Armitage} P.~J.,  {Livio} M.,   {Pringle} J.~E.,  2001, \mn@doi [\mnras]
  {10.1046/j.1365-8711.2001.04356.x}, \href
  {https://ui.adsabs.harvard.edu/abs/2001MNRAS.324..705A} {324, 705}

\bibitem[\protect\citeauthoryear{{Bai} \& {Stone}}{{Bai} \&
  {Stone}}{2013}]{Bai2013}
{Bai} X.-N.,  {Stone} J.~M.,  2013, \mn@doi [\apj]
  {10.1088/0004-637X/769/1/76}, \href
  {https://ui.adsabs.harvard.edu/abs/2013ApJ...769...76B} {769, 76}

\bibitem[\protect\citeauthoryear{{Bai}, {Ye}, {Goodman}  \& {Yuan}}{{Bai}
  et~al.}{2016}]{Bai2016}
{Bai} X.-N.,  {Ye} J.,  {Goodman} J.,   {Yuan} F.,  2016, \mn@doi [\apj]
  {10.3847/0004-637X/818/2/152}, \href
  {https://ui.adsabs.harvard.edu/abs/2016ApJ...818..152B} {818, 152}

\bibitem[\protect\citeauthoryear{{Bergin} \& {Williams}}{{Bergin} \&
  {Williams}}{2018}]{Bergin2018}
{Bergin} E.~A.,  {Williams} J.~P.,  2018, arXiv e-prints, \href
  {https://ui.adsabs.harvard.edu/abs/2018arXiv180709631B} {p. arXiv:1807.09631}

\bibitem[\protect\citeauthoryear{{Bergin} et~al.,}{{Bergin}
  et~al.}{2013}]{Bergin2013}
{Bergin} E.~A.,  et~al., 2013, \mn@doi [\nat] {10.1038/nature11805}, \href
  {https://ui.adsabs.harvard.edu/abs/2013Natur.493..644B} {493, 644}

\bibitem[\protect\citeauthoryear{{Birnstiel}, {Dullemond}  \&
  {Brauer}}{{Birnstiel} et~al.}{2010}]{birnstiel10}
{Birnstiel} T.,  {Dullemond} C.~P.,   {Brauer} F.,  2010, \mn@doi [\aap]
  {10.1051/0004-6361/200913731}, \href
  {http://adsabs.harvard.edu/abs/2010A...513A..79B} {513, A79}

\bibitem[\protect\citeauthoryear{{Birnstiel}, {Andrews}  \&
  {Ercolano}}{{Birnstiel} et~al.}{2012}]{birnstiel12}
{Birnstiel} T.,  {Andrews} S.~M.,   {Ercolano} B.,  2012, \mn@doi [\aap]
  {10.1051/0004-6361/201219262}, \href
  {http://adsabs.harvard.edu/abs/2012A%26A...544A..79B} {544, A79}

\bibitem[\protect\citeauthoryear{{Bitsch}, {Crida}, {Morbidelli}, {Kley}  \&
  {Dobbs-Dixon}}{{Bitsch} et~al.}{2013}]{Bitsch2013}
{Bitsch} B.,  {Crida} A.,  {Morbidelli} A.,  {Kley} W.,   {Dobbs-Dixon} I.,
  2013, \mn@doi [\aap] {10.1051/0004-6361/201220159}, \href
  {https://ui.adsabs.harvard.edu/abs/2013A&A...549A.124B} {549, A124}

\bibitem[\protect\citeauthoryear{{Booth} \& {Ilee}}{{Booth} \&
  {Ilee}}{2019}]{Booth2019}
{Booth} R.~A.,  {Ilee} J.~D.,  2019, \mn@doi [\mnras] {10.1093/mnras/stz1488},
  \href {https://ui.adsabs.harvard.edu/abs/2019MNRAS.487.3998B} {487, 3998}

\bibitem[\protect\citeauthoryear{{Booth} \& {Owen}}{{Booth} \&
  {Owen}}{2020}]{Booth2020}
{Booth} R.~A.,  {Owen} J.~E.,  2020, \mn@doi [\mnras] {10.1093/mnras/staa578},
  \href {https://ui.adsabs.harvard.edu/abs/2020MNRAS.493.5079B} {493, 5079}

\bibitem[\protect\citeauthoryear{{Booth}, {Clarke}, {Madhusudhan}  \&
  {Ilee}}{{Booth} et~al.}{2017}]{Booth2017}
{Booth} R.~A.,  {Clarke} C.~J.,  {Madhusudhan} N.,   {Ilee} J.~D.,  2017,
  \mn@doi [\mnras] {10.1093/mnras/stx1103}, \href
  {https://ui.adsabs.harvard.edu/abs/2017MNRAS.469.3994B} {469, 3994}

\bibitem[\protect\citeauthoryear{{Brauer}, {Dullemond}  \& {Henning}}{{Brauer}
  et~al.}{2008}]{Brauer2008}
{Brauer} F.,  {Dullemond} C.~P.,   {Henning} T.,  2008, \mn@doi [\aap]
  {10.1051/0004-6361:20077759}, \href
  {https://ui.adsabs.harvard.edu/abs/2008A&A...480..859B} {480, 859}

\bibitem[\protect\citeauthoryear{{Cannizzo}}{{Cannizzo}}{1993}]{Cannizzo1993}
{Cannizzo} J.~K.,  1993, \mn@doi [\apj] {10.1086/173486}, \href
  {https://ui.adsabs.harvard.edu/abs/1993ApJ...419..318C} {419, 318}

\bibitem[\protect\citeauthoryear{{Chiang} \& {Goldreich}}{{Chiang} \&
  {Goldreich}}{1997}]{chiang97}
{Chiang} E.~I.,  {Goldreich} P.,  1997, \mn@doi [\apj] {10.1086/304869}, \href
  {http://adsabs.harvard.edu/abs/1997ApJ...490..368C} {490, 368}

\bibitem[\protect\citeauthoryear{{Chiang}, {Joung}, {Creech-Eakman}, {Qi},
  {Kessler}, {Blake}  \& {van Dishoeck}}{{Chiang} et~al.}{2001}]{Chiang2001}
{Chiang} E.~I.,  {Joung} M.~K.,  {Creech-Eakman} M.~J.,  {Qi} C.,  {Kessler}
  J.~E.,  {Blake} G.~A.,   {van Dishoeck} E.~F.,  2001, \mn@doi [\apj]
  {10.1086/318427}, \href
  {https://ui.adsabs.harvard.edu/abs/2001ApJ...547.1077C} {547, 1077}

\bibitem[\protect\citeauthoryear{{Coleman}, {Kotko}, {Blaes}, {Lasota}  \&
  {Hirose}}{{Coleman} et~al.}{2016}]{Coleman2016}
{Coleman} M.~S.~B.,  {Kotko} I.,  {Blaes} O.,  {Lasota} J.~P.,   {Hirose} S.,
  2016, \mn@doi [\mnras] {10.1093/mnras/stw1908}, \href
  {https://ui.adsabs.harvard.edu/abs/2016MNRAS.462.3710C} {462, 3710}

\bibitem[\protect\citeauthoryear{{Cuzzi} \& {Zahnle}}{{Cuzzi} \&
  {Zahnle}}{2004}]{Cuzzi2004}
{Cuzzi} J.~N.,  {Zahnle} K.~J.,  2004, \mn@doi [\apj] {10.1086/423611}, \href
  {https://ui.adsabs.harvard.edu/abs/2004ApJ...614..490C} {614, 490}

\bibitem[\protect\citeauthoryear{{D'Alessio}, {Cant{\"o}}, {Calvet}  \&
  {Lizano}}{{D'Alessio} et~al.}{1998}]{dalessio98}
{D'Alessio} P.,  {Cant{\"o}} J.,  {Calvet} N.,   {Lizano} S.,  1998, \mn@doi
  [\apj] {10.1086/305702}, \href
  {https://ui.adsabs.harvard.edu/abs/1998ApJ...500..411D} {500, 411}

\bibitem[\protect\citeauthoryear{{D'Alessio}, {Calvet}  \&
  {Hartmann}}{{D'Alessio} et~al.}{2001}]{dalessio01}
{D'Alessio} P.,  {Calvet} N.,   {Hartmann} L.,  2001, \mn@doi [\apj]
  {10.1086/320655}, \href {http://adsabs.harvard.edu/abs/2001ApJ...553..321D}
  {553, 321}

\bibitem[\protect\citeauthoryear{{Dullemond}}{{Dullemond}}{2000}]{Dullemond2000}
{Dullemond} C.~P.,  2000, \aap, \href
  {https://ui.adsabs.harvard.edu/abs/2000A&A...361L..17D} {361, L17}

\bibitem[\protect\citeauthoryear{{Dullemond} \& {Natta}}{{Dullemond} \&
  {Natta}}{2003}]{Dullemond2003}
{Dullemond} C.~P.,  {Natta} A.,  2003, \mn@doi [\aap]
  {10.1051/0004-6361:20030606}, \href
  {https://ui.adsabs.harvard.edu/abs/2003A&A...405..597D} {405, 597}

\bibitem[\protect\citeauthoryear{{Dullemond}, {Dominik}  \&
  {Natta}}{{Dullemond} et~al.}{2001}]{Dullemond2001}
{Dullemond} C.~P.,  {Dominik} C.,   {Natta} A.,  2001, \mn@doi [\apj]
  {10.1086/323057}, \href
  {https://ui.adsabs.harvard.edu/abs/2001ApJ...560..957D} {560, 957}

\bibitem[\protect\citeauthoryear{{Faulkner}, {Lin}  \& {Papaloizou}}{{Faulkner}
  et~al.}{1983}]{Faulkner1983}
{Faulkner} J.,  {Lin} D.~N.~C.,   {Papaloizou} J.,  1983, \mn@doi [\mnras]
  {10.1093/mnras/205.2.359}, \href
  {https://ui.adsabs.harvard.edu/abs/1983MNRAS.205..359F} {205, 359}

\bibitem[\protect\citeauthoryear{{Fedele} et~al.,}{{Fedele}
  et~al.}{2018}]{Fedele2018}
{Fedele} D.,  et~al., 2018, \mn@doi [\aap] {10.1051/0004-6361/201731978}, \href
  {https://ui.adsabs.harvard.edu/abs/2018A&A...610A..24F} {610, A24}

\bibitem[\protect\citeauthoryear{{Gressel}, {Turner}, {Nelson}  \&
  {McNally}}{{Gressel} et~al.}{2015}]{Gressel2015}
{Gressel} O.,  {Turner} N.~J.,  {Nelson} R.~P.,   {McNally} C.~P.,  2015,
  \mn@doi [\apj] {10.1088/0004-637X/801/2/84}, \href
  {https://ui.adsabs.harvard.edu/abs/2015ApJ...801...84G} {801, 84}

\bibitem[\protect\citeauthoryear{{Guidi} et~al.,}{{Guidi}
  et~al.}{2016}]{Guidi2016}
{Guidi} G.,  et~al., 2016, \mn@doi [\aap] {10.1051/0004-6361/201527516}, \href
  {https://ui.adsabs.harvard.edu/abs/2016A&A...588A.112G} {588, A112}

\bibitem[\protect\citeauthoryear{{Hayashi}}{{Hayashi}}{1981}]{Hayashi1981}
{Hayashi} C.,  1981, \mn@doi [Progress of Theoretical Physics Supplement]
  {10.1143/PTPS.70.35}, \href
  {https://ui.adsabs.harvard.edu/abs/1981PThPS..70...35H} {70, 35}

\bibitem[\protect\citeauthoryear{Hindmarsh}{Hindmarsh}{1983}]{Hindmarsh1983}
Hindmarsh A.~C.,  1983, Scientific Computing, pp 55--64

\bibitem[\protect\citeauthoryear{{Hollenbach}, {Kaufman}, {Bergin}  \&
  {Melnick}}{{Hollenbach} et~al.}{2009}]{Hollenbach2009}
{Hollenbach} D.,  {Kaufman} M.~J.,  {Bergin} E.~A.,   {Melnick} G.~J.,  2009,
  \mn@doi [\apj] {10.1088/0004-637X/690/2/1497}, \href
  {https://ui.adsabs.harvard.edu/abs/2009ApJ...690.1497H} {690, 1497}

\bibitem[\protect\citeauthoryear{{Huang} et~al.,}{{Huang}
  et~al.}{2018}]{Huang2018}
{Huang} J.,  et~al., 2018, \mn@doi [\apjl] {10.3847/2041-8213/aaf740}, \href
  {https://ui.adsabs.harvard.edu/abs/2018ApJ...869L..42H} {869, L42}

\bibitem[\protect\citeauthoryear{{Isella} et~al.,}{{Isella}
  et~al.}{2016}]{Isella2016}
{Isella} A.,  et~al., 2016, \mn@doi [\prl] {10.1103/PhysRevLett.117.251101},
  \href {https://ui.adsabs.harvard.edu/abs/2016PhRvL.117y1101I} {117, 251101}

\bibitem[\protect\citeauthoryear{{Kama} et~al.,}{{Kama}
  et~al.}{2020}]{Kama2020}
{Kama} M.,  et~al., 2020, \mn@doi [\aap] {10.1051/0004-6361/201937124}, \href
  {https://ui.adsabs.harvard.edu/abs/2020A&A...634A..88K} {634, A88}

\bibitem[\protect\citeauthoryear{{Kretke} \& {Lin}}{{Kretke} \&
  {Lin}}{2007}]{Kretke2007}
{Kretke} K.~A.,  {Lin} D.~N.~C.,  2007, \mn@doi [\apjl] {10.1086/520718}, \href
  {https://ui.adsabs.harvard.edu/abs/2007ApJ...664L..55K} {664, L55}

\bibitem[\protect\citeauthoryear{{Krijt}, {Schwarz}, {Bergin}  \&
  {Ciesla}}{{Krijt} et~al.}{2018}]{Krijt2018}
{Krijt} S.,  {Schwarz} K.~R.,  {Bergin} E.~A.,   {Ciesla} F.~J.,  2018, \mn@doi
  [\apj] {10.3847/1538-4357/aad69b}, \href
  {https://ui.adsabs.harvard.edu/abs/2018ApJ...864...78K} {864, 78}

\bibitem[\protect\citeauthoryear{{Kuiper}, {Klahr}, {Dullemond}, {Kley}  \&
  {Henning}}{{Kuiper} et~al.}{2010}]{Kuiper2010}
{Kuiper} R.,  {Klahr} H.,  {Dullemond} C.,  {Kley} W.,   {Henning} T.,  2010,
  \mn@doi [\aap] {10.1051/0004-6361/200912355}, \href
  {https://ui.adsabs.harvard.edu/abs/2010A&A...511A..81K} {511, A81}

\bibitem[\protect\citeauthoryear{{Lasota}}{{Lasota}}{2001}]{Lasota2001}
{Lasota} J.-P.,  2001, \mn@doi [\nar] {10.1016/S1387-6473(01)00112-9}, \href
  {https://ui.adsabs.harvard.edu/abs/2001NewAR..45..449L} {45, 449}

\bibitem[\protect\citeauthoryear{{Madhusudhan}, {Amin}  \&
  {Kennedy}}{{Madhusudhan} et~al.}{2014}]{Madhusudhan2014}
{Madhusudhan} N.,  {Amin} M.~A.,   {Kennedy} G.~M.,  2014, \mn@doi [\apjl]
  {10.1088/2041-8205/794/1/L12}, \href
  {https://ui.adsabs.harvard.edu/abs/2014ApJ...794L..12M} {794, L12}

\bibitem[\protect\citeauthoryear{{Martin} \& {Livio}}{{Martin} \&
  {Livio}}{2012}]{Martin2012}
{Martin} R.~G.,  {Livio} M.,  2012, \mn@doi [\mnras]
  {10.1111/j.1745-3933.2012.01290.x}, \href
  {https://ui.adsabs.harvard.edu/abs/2012MNRAS.425L...6M} {425, L6}

\bibitem[\protect\citeauthoryear{{Martin} \& {Livio}}{{Martin} \&
  {Livio}}{2014}]{Martin2014}
{Martin} R.~G.,  {Livio} M.,  2014, \mn@doi [\apjl]
  {10.1088/2041-8205/783/2/L28}, \href
  {https://ui.adsabs.harvard.edu/abs/2014ApJ...783L..28M} {783, L28}

\bibitem[\protect\citeauthoryear{{Martin} \& {Lubow}}{{Martin} \&
  {Lubow}}{2011}]{Martin2011}
{Martin} R.~G.,  {Lubow} S.~H.,  2011, \mn@doi [\apjl]
  {10.1088/2041-8205/740/1/L6}, \href
  {https://ui.adsabs.harvard.edu/abs/2011ApJ...740L...6M} {740, L6}

\bibitem[\protect\citeauthoryear{{Martin} \& {Lubow}}{{Martin} \&
  {Lubow}}{2013}]{Martin2013}
{Martin} R.~G.,  {Lubow} S.~H.,  2013, \mn@doi [\mnras] {10.1093/mnras/stt580},
  \href {https://ui.adsabs.harvard.edu/abs/2013MNRAS.432.1616M} {432, 1616}

\bibitem[\protect\citeauthoryear{{McClure} et~al.,}{{McClure}
  et~al.}{2016}]{McClure2016}
{McClure} M.~K.,  et~al., 2016, \mn@doi [\apj] {10.3847/0004-637X/831/2/167},
  \href {https://ui.adsabs.harvard.edu/abs/2016ApJ...831..167M} {831, 167}

\bibitem[\protect\citeauthoryear{{Min}, {Dullemond}, {Kama}  \&
  {Dominik}}{{Min} et~al.}{2011}]{Min2011}
{Min} M.,  {Dullemond} C.~P.,  {Kama} M.,   {Dominik} C.,  2011, \mn@doi
  [\icarus] {10.1016/j.icarus.2010.12.002}, \href
  {https://ui.adsabs.harvard.edu/abs/2011Icar..212..416M} {212, 416}

\bibitem[\protect\citeauthoryear{{{\"O}berg}, {Murray-Clay}  \&
  {Bergin}}{{{\"O}berg} et~al.}{2011}]{Oberg2011}
{{\"O}berg} K.~I.,  {Murray-Clay} R.,   {Bergin} E.~A.,  2011, \mn@doi [\apjl]
  {10.1088/2041-8205/743/1/L16}, \href
  {https://ui.adsabs.harvard.edu/abs/2011ApJ...743L..16O} {743, L16}

\bibitem[\protect\citeauthoryear{{Okuzumi}, {Momose}, {Sirono}, {Kobayashi}  \&
  {Tanaka}}{{Okuzumi} et~al.}{2016}]{Okuzumi2016}
{Okuzumi} S.,  {Momose} M.,  {Sirono} S.-i.,  {Kobayashi} H.,   {Tanaka} H.,
  2016, \mn@doi [\apj] {10.3847/0004-637X/821/2/82}, \href
  {https://ui.adsabs.harvard.edu/abs/2016ApJ...821...82O} {821, 82}

\bibitem[\protect\citeauthoryear{{Owen}}{{Owen}}{2014}]{Owen2014b}
{Owen} J.~E.,  2014, \mn@doi [\apj] {10.1088/0004-637X/789/1/59}, \href
  {https://ui.adsabs.harvard.edu/abs/2014ApJ...789...59O} {789, 59}

\bibitem[\protect\citeauthoryear{{Owen} \& {Armitage}}{{Owen} \&
  {Armitage}}{2014}]{Owen2014}
{Owen} J.~E.,  {Armitage} P.~J.,  2014, \mn@doi [\mnras]
  {10.1093/mnras/stu1928}, \href
  {https://ui.adsabs.harvard.edu/abs/2014MNRAS.445.2800O} {445, 2800}

\bibitem[\protect\citeauthoryear{{Piso}, {{\"O}berg}, {Birnstiel}  \&
  {Murray-Clay}}{{Piso} et~al.}{2015}]{Piso2015}
{Piso} A.-M.~A.,  {{\"O}berg} K.~I.,  {Birnstiel} T.,   {Murray-Clay} R.~A.,
  2015, \mn@doi [\apj] {10.1088/0004-637X/815/2/109}, \href
  {https://ui.adsabs.harvard.edu/abs/2015ApJ...815..109P} {815, 109}

\bibitem[\protect\citeauthoryear{{Powell}, {Murray-Clay}  \&
  {Schlichting}}{{Powell} et~al.}{2017}]{Powell2017}
{Powell} D.,  {Murray-Clay} R.,   {Schlichting} H.~E.,  2017, \mn@doi [\apj]
  {10.3847/1538-4357/aa6d7c}, \href
  {https://ui.adsabs.harvard.edu/abs/2017ApJ...840...93P} {840, 93}

\bibitem[\protect\citeauthoryear{{Pringle}}{{Pringle}}{1981}]{Pringle1981}
{Pringle} J.~E.,  1981, \mn@doi [\araa] {10.1146/annurev.aa.19.090181.001033},
  \href {https://ui.adsabs.harvard.edu/abs/1981ARA&A..19..137P} {19, 137}

\bibitem[\protect\citeauthoryear{{Qi} et~al.,}{{Qi} et~al.}{2013}]{Qi_science}
{Qi} C.,  et~al., 2013, \mn@doi [Science] {10.1126/science.1239560}, \href
  {http://adsabs.harvard.edu/abs/2013Sci...341..630Q} {341, 630}

\bibitem[\protect\citeauthoryear{{Qi} et~al.,}{{Qi} et~al.}{2019}]{Qi2019}
{Qi} C.,  et~al., 2019, \mn@doi [\apj] {10.3847/1538-4357/ab35d3}, \href
  {https://ui.adsabs.harvard.edu/abs/2019ApJ...882..160Q} {882, 160}

\bibitem[\protect\citeauthoryear{{Rafikov} \& {De Colle}}{{Rafikov} \& {De
  Colle}}{2006}]{Rafikov2006}
{Rafikov} R.~R.,  {De Colle} F.,  2006, \mn@doi [\apj] {10.1086/504833}, \href
  {https://ui.adsabs.harvard.edu/abs/2006ApJ...646..275R} {646, 275}

\bibitem[\protect\citeauthoryear{{Schwarz}, {Bergin}, {Cleeves}, {Zhang},
  {{\"O}berg}, {Blake}  \& {Anderson}}{{Schwarz} et~al.}{2018}]{Schwarz2018}
{Schwarz} K.~R.,  {Bergin} E.~A.,  {Cleeves} L.~I.,  {Zhang} K.,  {{\"O}berg}
  K.~I.,  {Blake} G.~A.,   {Anderson} D.,  2018, \mn@doi [\apj]
  {10.3847/1538-4357/aaae08}, \href
  {https://ui.adsabs.harvard.edu/abs/2018ApJ...856...85S} {856, 85}

\bibitem[\protect\citeauthoryear{{Stammler}, {Birnstiel}, {Pani{\'c}},
  {Dullemond}  \& {Dominik}}{{Stammler} et~al.}{2017}]{Stammler2017}
{Stammler} S.~M.,  {Birnstiel} T.,  {Pani{\'c}} O.,  {Dullemond} C.~P.,
  {Dominik} C.,  2017, \mn@doi [\aap] {10.1051/0004-6361/201629041}, \href
  {https://ui.adsabs.harvard.edu/abs/2017A&A...600A.140S} {600, A140}

\bibitem[\protect\citeauthoryear{Stephenson \& Malanowski}{Stephenson \&
  Malanowski}{1987}]{REF:NIST}
Stephenson R.~M.,  Malanowski S.,  1987, {Handbook of the thermodynamics of
  organic compounds}.
Springer, Dordrecht, \mn@doi{10.1007/978-94-009-3173-2}, \url
  {http://cds.cern.ch/record/2006502}

\bibitem[\protect\citeauthoryear{{Stevenson} \& {Lunine}}{{Stevenson} \&
  {Lunine}}{1988}]{Stevenson1988}
{Stevenson} D.~J.,  {Lunine} J.~I.,  1988, \mn@doi [\icarus]
  {10.1016/0019-1035(88)90133-9}, \href
  {https://ui.adsabs.harvard.edu/abs/1988Icar...75..146S} {75, 146}

\bibitem[\protect\citeauthoryear{{Takeuchi} \& {Lin}}{{Takeuchi} \&
  {Lin}}{2002}]{takeuchi02}
{Takeuchi} T.,  {Lin} D.~N.~C.,  2002, \mn@doi [\apj] {10.1086/344437}, \href
  {http://adsabs.harvard.edu/abs/2002ApJ...581.1344T} {581, 1344}

\bibitem[\protect\citeauthoryear{{Tinetti} et~al.,}{{Tinetti}
  et~al.}{2018}]{Tinetti2018}
{Tinetti} G.,  et~al., 2018, \mn@doi [Experimental Astronomy]
  {10.1007/s10686-018-9598-x}, \href
  {https://ui.adsabs.harvard.edu/abs/2018ExA....46..135T} {46, 135}

\bibitem[\protect\citeauthoryear{{Williams} \& {Best}}{{Williams} \&
  {Best}}{2014}]{Williams2014}
{Williams} J.~P.,  {Best} W. M.~J.,  2014, \mn@doi [\apj]
  {10.1088/0004-637X/788/1/59}, \href
  {https://ui.adsabs.harvard.edu/abs/2014ApJ...788...59W} {788, 59}

\bibitem[\protect\citeauthoryear{{Zhang}, {Blake}  \& {Bergin}}{{Zhang}
  et~al.}{2015}]{Zhang2015}
{Zhang} K.,  {Blake} G.~A.,   {Bergin} E.~A.,  2015, \mn@doi [\apjl]
  {10.1088/2041-8205/806/1/L7}, \href
  {https://ui.adsabs.harvard.edu/abs/2015ApJ...806L...7Z} {806, L7}

\bibitem[\protect\citeauthoryear{{Zhang} et~al.,}{{Zhang}
  et~al.}{2018}]{Zhang2018}
{Zhang} S.,  et~al., 2018, \mn@doi [\apjl] {10.3847/2041-8213/aaf744}, \href
  {https://ui.adsabs.harvard.edu/abs/2018ApJ...869L..47Z} {869, L47}

\bibitem[\protect\citeauthoryear{{Zhu}, {Hartmann}, {Gammie}, {Book}, {Simon}
  \& {Engelhard}}{{Zhu} et~al.}{2010}]{Zhu2010}
{Zhu} Z.,  {Hartmann} L.,  {Gammie} C.~F.,  {Book} L.~G.,  {Simon} J.~B.,
  {Engelhard} E.,  2010, \mn@doi [\apj] {10.1088/0004-637X/713/2/1134}, \href
  {https://ui.adsabs.harvard.edu/abs/2010ApJ...713.1134Z} {713, 1134}

\bibitem[\protect\citeauthoryear{{van der Marel}, {Dong}, {di Francesco},
  {Williams}  \& {Tobin}}{{van der Marel} et~al.}{2019}]{vanderMarel2019}
{van der Marel} N.,  {Dong} R.,  {di Francesco} J.,  {Williams} J.~P.,
  {Tobin} J.,  2019, \mn@doi [\apj] {10.3847/1538-4357/aafd31}, \href
  {https://ui.adsabs.harvard.edu/abs/2019ApJ...872..112V} {872, 112}

\makeatother
\end{thebibliography}

%%%%%%%%%%%%%%%%%%%%%%%%%%%%%%%%%%%%%%%%%%%%%%%%%%

% Don't change these lines
\bsp	% typesetting comment
\label{lastpage}
\end{document}